# Causal Sets from simple models of computation

**Tommaso Bolognesi**
CNR/ISTI - Via Moruzzi 1, 56124 Pisa, Italy
t.bolognesi@isti.cnr.it

**Abstract**

Causality among events is widely recognized as a most fundamental structure of spacetime, and causal sets have been proposed as discrete models of the latter in the context of quantum gravity theories, notably in the Causal Set Programme.

In the rather different context of what might be called the 'Computational Universe Programme' -- one which associates the complexity of physical phenomena to the emergent features of models such as cellular automata -- a choice problem arises with respect to the variety of formal systems that, in virtue of their computational universality (Turing-completeness), qualify as equally good candidates for a computational, unified theory of physics.

This paper proposes Causal Sets as the only objects of physical significance and relevance to be considered under the 'computational universe' perspective, and as the appropriate abstraction for shielding the unessential details of the many different computationally universal candidate models. At the same time, we propose a fully deterministic, radical alternative to the probabilistic techniques currently considered in the Causal Set Programme for growing discrete spacetimes.

We investigate a number of computation models by grouping them into two broad classes, based on the support on which they operate; in one case this is linear, like a tape or a string of symbols; in the other, it is a two-dimensional grid or a planar graph. For each model we identify the causality relation among computation events, implement it, and conduct a possibly exhaustive exploration of the associated causal set space, while examining quantitative and qualitative features such as dimensionality, curvature, planarity, emergence of pseudo-randomness, causal set substructures and particles.

## 1. Introduction

In the last few decades several scientists (K. Zuse, J. A. Wheeler, R. Feynman, E. Fredkin, S. Wolfram, G. 't Hooft, S. Lloyd, J. Schmidhuber, to mention a few) have contributed, in a variety of ways and degrees, to the birth and progression of a wave of ideas and conjectures that could be collectively named 'Computational Universe Programme'. In its most extreme form, this programme suggests that all the complexity we observe in the physical universe might emerge from the iteration of a few simple transition rules, that could be implemented by a short computer program. The fact that simple programs can produce highly complex patterns, sometimes similar to those found in nature, is widely recognized today, and has been investigated and divulgated, in particular, by S. Wolfram, with his extensive behavioural analysis of cellular automata and other simple models [21]. These facts are taken by some researchers -- not without considerable skepticism by others -- as a basis, or, at least, an inspiring metaphor for trying to devise a radically new, computation-based, spectacularly simple fundamental theory of physics.

Once the idea is adopted of understanding complexity in physics as emergence in computation, an obvious question arises: which model of computation? It seems reasonable to restrict to 'universal' models, in the sense of Turing-complete, since computational universality is indeed supported by our physical universe (e.g. in computers!). But universality turns out to be quite cheap, and at reach for many simple formal models; according to a conjecture by Wolfram [21], universality is indeed found in any artificial or natural system capable of pseudo-random behaviour.

Fredkin [4] suggests an interesting choice criterion:

*"We might be able to demonstrate that an ordinary computer model of physics is sufficient, but we cannot normally show that it is necessary. The reason is that any and all models of finite nature can be replaced by equivalent computational models based on any universal computer. [...] DM (Digital Mechanics) implies that there is a computer-like model that has a bijective mapping, one to one, from states and function in the real world to states and function in the model. [...] The beauty of the 'one-to-one mapping onto' restriction is that it delivers us from the apparent tyranny of computation universality. It is unlikely that there will be more than one such correct model."*

An illustration of the above concept is provided by Fredkin himself: a flight simulator running on a PC is not a good model of an airplane since many of the data items manipulated by the computer are not related to the airplane, but, say, to the PC operating system, and, conversely, many structural and behavioral aspects of the real airplane are abstracted away in the simulation.

Based on these criteria, Fredkin is lead to focus on a precise family of computation models, namely cellular automata; in particular, he investigates second-order, reversible, universal automata (RUCA), and a conservative model called SALT [10].

In this paper we retain Fredkin's general requirement of a bijective mapping, but provide an alternative, somewhat less restrictive formulation of it, through the notion of *causal set*. Our approach is mainly motivated by two observations:
- Devising partial orders and causal relations among computation events appears to be feasible for a number of formal models, although, ironically, this is not the case for cellular automata -- probably the most frequently considered model of the Computational Universe Programme.
- Causality among spacetime events is regarded as a most fundamental aspect of nature, and, when combining this notion with spacetime discreteness at the Plank scale -- an assumption commonly made also under the computational universe view -- the notion of causal set inevitably comes on stage, as it happens with the Causal Set approach to quantum gravity [3, 14, 15, 16, 18].

Our idea is to understand Fredkin's one-to-one mapping between the computational model and physical reality exclusively as a correspondence between formal and physical *events*, without promoting to the status of 'reality' any component of the underlying data structure, or *state*. This is motivated by considering that any observation process in which we might engage ultimately consists in a structure of causally related events in spacetime. Thus, by claiming that some Turing machine correctly models -- or precisely computes -- our physical universe, we would only refer to the set of atomic events forming its computation, and to their causal relationships, not to the machine control head states or tape configurations.

Following the idea of abstracting from the underlying state structure, one might be tempted to formulate a bold conjecture, asserting the equivalence of all universal models of computation not only in the context of computer science but also for describing the physical universe: each known universal model would admit a specific computation, with appropriate initial conditions, that would yield the same causal set, and the *correct* one for our physical universe. Such a result would provide a clean answer to the 'tyranny of computational universality' problem, somehow opposite to Fredkin's solution: *all* universal models can compute our history, and there is no need to choose one in particular. However the differences we have detected among causal sets from different models, even regardless of initial conditions, seem to leave very little room for this conjecture, as we shall see in the subsequent sections.

The notion of causality among spacetime events is of course central in relativity theory. In the flat, continuous spacetime of special relativity (Minkowski space) two events $(ct, x, y, z)$ and $(ct', x', y', z')$ -- where $t$ is the time coordinate, $x, y, z$ are the space coordinates, and $c$ is the speed of light -- are causally related when their spacetime distance

$$d^2 = c^2(t - t')^2 - (x - x')^2 - (y - y')^2 - (z - z')^2$$

under a (+ - - -) signature, is positive or null; a photon can travel along a *timelike* line carrying information, and influence, from one to the other. In general, if we are given the metric information $g_{\mu\nu}$ of a continuous spacetime, we use $ds^2 = g_{\mu\nu} dx^\mu dx^\nu$ for determining all possible causal relations among events, thus obtaining the familiar light-cone patterns.

Conversely, the causal structure alone provides us with all the information needed to determine the metric, and the gravitational field tensor, up to a multiplicative conformal factor. (For these remarks we follow [14], which in turn refers to [5, 7].) Without knowledge of this conformal factor, we miss the scale that enables us to come up with well defined measures of length and volume in spacetime. But when spacetime is discrete, and its causal structure is coded explicitly and completely in a directed graph -- the causal set --, then those measures can be achieved by essentially counting the nodes in a portion of the graph, and we have enough information for completely determining the metric tensors of general relativity. This is the reason why causal sets are regarded as the most fundamental structure of spacetime by the research programme named after them.

A *causal set* ('*causet*') is a *finitary* (or 'locally finite') *partially ordered set*, that is, a set of events with a partial order relation '<' which is:

1. *reflexive*: for any event $x$, $x<x$.
2. *anti-symmetric*: if $x<y$ and $y<x$ then $x=y$;
3. *transitive*: if $x<y$ and $y<z$, then $x<z$;
4. *finitarity*: the number of events between any two events is finite.

Causal sets can be drawn as directed, acyclic graphs. For convenience, what is usually shown is the *transitive reduction* of these graphs, also called *Hasse diagram*, obtained by removing all 'redundant' arcs -- those implied by transitivity; the complete graph can be recovered by *transitive closure*.

Several of the papers from the contributors to the Causal Set Programme emphasize on the concept of 'sprinkling', which refers to the process of creating a causal set $C(M)$ by generating points in a manifold $M$ with Lorentzian metric according to a Poisson distribution of unit density. The discrete structure inherit causality from the continuous one, and the nature of the distribution guarantees that the number of points found in a region of $M$ is proportional to the volume, thus justifying the equation "geometry = order + *number*". Sprinkling and *coarse graining* (see the referenced papers for details) bridge the gap between the causet and the Lorentzian manifold -- the latter intended as a macroscopic, continuous approximation of the former. Then, whenever a *direct* procedure is devised for growing some causet $C$, one may inquire about the 'manifoldness' of $C$, and check whether it might have been produced by a sprinkling of a spacetime manifold $M$.

Our main goals are:

- to rigorously define procedures for deriving causets from simple models of computation;
- to investigate the classes of causets that can be thus obtained, their emergent features, and their potential commonalities.

Given that the territory we are entering here is largely unexplored, our attitude has been one of avoiding specific expectations, while being basically open to *anything interesting* that might emerge. In doing so we have been in part inspired by the experimental and largely visual investigation style of [21]. Furthermore, in our analysis of the causal set spaces associated with various models of computation we have definitely taken advantage of the massive, experimental and classificatory work carried out in [21], which has led to the selection of the most interesting instances for some of those models. Our study has progressively lead us to focus on a number of features, namely planarity, dimensionality, curvature, and emergence of pseudo-randomness, causal set substructures and particles: these aspects have been given higher priority over the 'Lorentzian manifoldness' issue, whose study is deferred to a later stage.

The rest of the paper is organized as follows.

In Section 2 we introduce the general idea of associating a causal set (*causet*, for short) to a computation, and illustrate it by considering the special case of Turing machines. Turing machine causets are provably planar, and can yield euclidean as well as hyperbolic geometries. We explore machine classes of growing complexity, analyze their dimensionality and curvature, show how the pseudo-random behavior of some computation is reflected in the causet structure, and contrast the causet representation with the 'compression technique' discussed in [21]. In Section 3 we consider four models that compute on a linear support -- a tape, or a string of symbols -- and show that for all of them the derived causal set is planar, thus generalizing the result for Turing machines. In Section 4 we consider two more models of computation: Turing machines on two-dimensional square grids, and mobile automata that operate on planar, trivalent graphs, observing increased causet dimensions, a form of 'particle', and a pseudo-random causet with signs of emergent causal substructure. In Section 5 we summarize and discuss our results, and mention some items for future work.

# 2. Causal sets from computations: the case of Turing machines

In [21, Ch. 9], Wolfram introduces a procedure for deriving causal sets (called there 'causal networks') from mobile automata on strings, and from string rewrite systems, suggesting that they could represent a structure for spacetime. We shall deal with these models of computation in the next section. Here we describe the general method for deriving a causet *CS* from a computation *C* of some formal model *M*, and then illustrate it for the case of Turing machines (TM). A demonstration on building the causal network of a specific TM has been developed by Wolfram and Nochella [22]. We view computation *C* as a sequence:

$$C = S_0, e_1, S_1, e_2, ..., S_{n-1}, e_n, S_n, ...$$

in which global states (the $S_i$'s) and events (the $e_i$'s) alternate; $S_0$ is the initial state. The global state structure depends on the model of computation *M*, and may involve various types of data variables, while events are unstructured and unlabeled, so that event $e_i$ could as well be identified by the pair $(S_{i-1}, S_i)$ of states. (An exception is represented by interactive models such as process algebras, in which events are themselves structured and/or labeled, and used for interaction with the external environment, but in this paper we only deal with 'closed' models of computation.) Then, every event in *C* becomes an event (a node) in causet *CS*, and a causal link is created in *CS* from $e_i$ to $e_j$, with $i < j$, whenever $e_i$ affects some component $x$ of the global state, and $e_j$ is the first event occuring after $e_i$ to read or make use of that component. Causality between the two events is *induced*, or *mediated*, by component $x$.

The application of this method to Turing machines is straightforward. The global state structure is composed by (i) a tape of cells which contain symbols from a finite alphabet, (ii) a finite-state variable describing the state of the control unit, and (iii) an indication of the position of the control unit on the tape. Thus, a global state could be represented by a finite string such as

$$S_{n-1} = a_1, a_2, ..., (s, a_i), a_{i+1}, ..., a_m$$

describing a finite portion of the tape, with the control unit in state *s*, positioned on a cell containing symbol $a_i$; the infinite portion of the tape not explicitly described by the string is thought of as filled by blank symbols. Assuming that the state transition table for the machine contains the entry $(s, a_i) \to (s', a_j, +1)$, the next global state, after event $e_n$, would be:

$$S_n = a_1, a_2, ..., a_j, (s', a_{i+1}), ..., a_m$$

where symbol $a_i$ has been replaced by $a_j$ and the control head has moved one step to the right. Every event $e_n$ affects (writes) three components of the global state -- the control head state, the control head position, and the content of a cell -- and for doing so it must also access (read) all three. However, while the first two components are always determined by the preceding event $e_{n-1}$, the cell content must have been determined earlier than that, since at every step the control head always moves to the left or to the right. Thus, for the generic event $e_n$ we always create a causal link $e_{n-1} \to e_n$, induced by the change of control head state and position, and another link $e_k \to e_n$, with $k < n-1$, whenever $e_k$ is be most recent event that has updated the cell read by $e_n$, or no link, when no such event exists. As a consequence, each node has at most two incoming and at most two outgoing arcs.

**Proposition 1.** *The causet of a Turing machine computation is planar.*

Proof. The planarity of the causet graph is established immediately if the computation is depicted by stacking the sequence of tape configurations as done in Figure 1. In that figure each event, represented as a circle, overlaps two successive configurations of the individual cell it reads and writes, thus reflecting the current position of the control head; causal links induced by cell read/write operations are depicted as vertical edges, and links induced by control head state and position are depicted as transversal edges. □

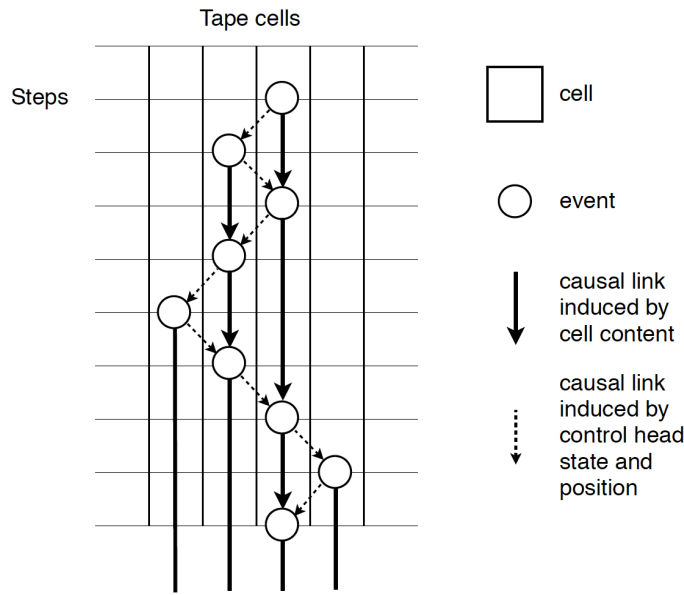

Figure 1 - Causet construction for a Turing machine computation.

Note that no information on actual cell content is directly coded in the causet. The presence of the causal links mediated by the control head state and position, that relate every event to its successor in the computation, has the remarkable consequence that the partial order is indeed total; the transitive reduction of the causet from any computation of any Turing machine always collapses to a linear path. Other models of computation involve similar 'single-step' causal links, as we shall see later, and share with Turing machines these totally uninteresting transitive reductions. For this reason we have chosen to attribute equal importance to *all* the causal links obtained by the criterion described above, in spite of the fact that, under a strict notion of causality, some of them are redundant. The choice of not excluding 'redundant' arcs yields a much richer scenario for our investigations, and has a direct impact, for example, on dimensionality analysis. But there is a less pragmatic justification for retaining these arcs: they are the only bridge between the causet -- intended as a spacetime -- and the *spatial* structure of the underlying support, which is not directly accessible at the spacetime level.

## 2.1 Causets for the class of elementary Turing machines: dimensionality and curvature

In an *elementary Turing* machine the control head has only two states, and the tape alphabet is binary. Each entry of the 2×2 state transition table is essentially a triple of bits -- one for the next state, one for the written symbol, and one for the left/right move, thus there are a total of $2^{3 \times 4} = 4096$ possible machines, whose computations we always conventionally start with the head in state 1 and the tape entirely filled with 0's. Out of 4096 causets, 4084 are linear, of the type illustrated in Figure 2.

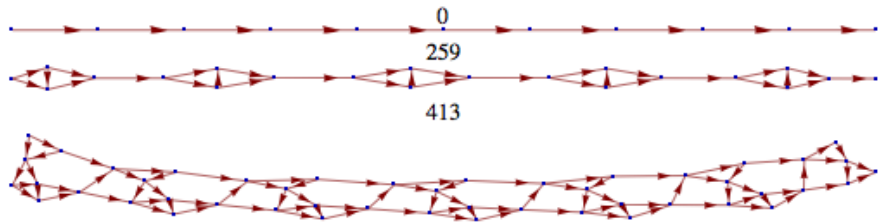

Figure 2 - Linear causets from computations of elementary Turing machines.

The remaining 12 causets are illustrated in Figure 3, where edge orientation has been omitted.

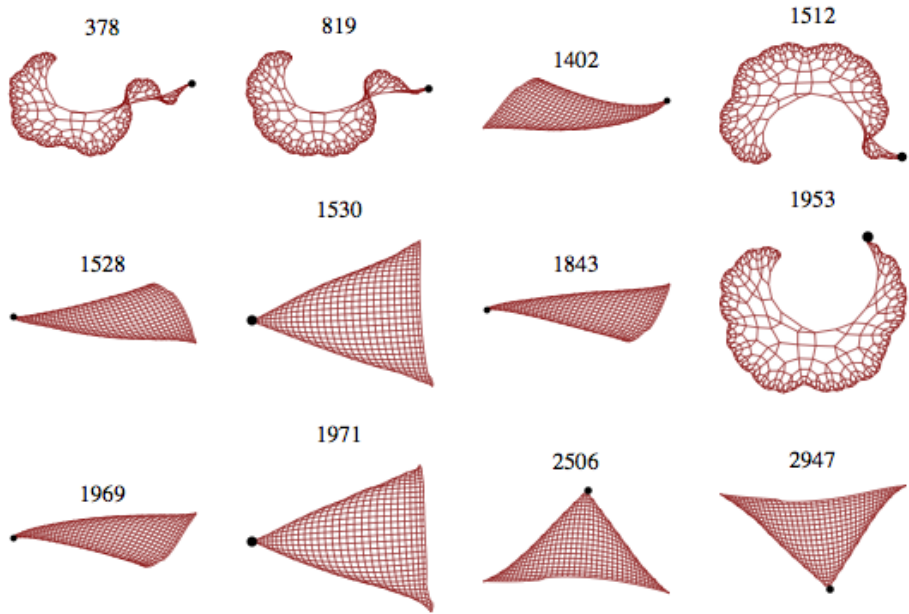

Figure 3 - The 8 Euclidean and 4 hyperbolic causets of elementary Turing machines.

The above machines can be grouped into six pairs, by left-right symmetry. For example, the transition table for TM 819 is obtained by inverting head moves in the table for TM 378:

Table for TM 378: {{1, 0} → {2, 0, 1}, {1, 1} → {1, 0, -1}, {2, 0} → {1, 1, -1}, {2, 1} → {2, 1, 1}};
Table for TM 819: {{1, 0} → {2, 0, -1}, {1, 1} → {1, 0, 1}, {2, 0} → {1, 1, 1}, {2, 1} → {2, 1, -1}}.

All the 4096 causets are, expectedly, planar, a circumstance that suggests a maximum value of 2 for their dimensionality. (In the sequel we shall use 'dimension', in place of 'dimensionality', with the understanding that we do not refer to set cardinality.) However, several dimension estimators have been defined for dealing with graphs, and some options are also available when addressing their curvature. We have chosen to use a well known graph theoretic dimension estimator based on detecting the growth rate of the size of node-shells found at progressive distance from a given node; in the case of *undirected* graphs this concept is defined, for example, in [11], where it is called *connectivity dimension*, and used in [21].

***Definition 1.*** *Neighborhood, surface.*

Let $C(N, E)$ be a causet, where $N$ and $E$ are the sets of nodes and edges, and let $x$ be a generic node. For any natural number $k$, the *k-neighborhood* of $x$ is the set $U_x(k) := \{n \in N \mid d(x, n) \leq k\}$, where $d$ is the distance from $x$ to $n$, intended as the length of the shortest path (counting edges) from $x$ to $n$. When no such path exists, then $d(x, n) := \infty$. Note that, with directed graphs, $d$ is not a distance in the usual sense, since it is not symmetric. The boundary, or *surface* of neighborhood $U_x(k)$ is the set $S_x(k) := U_x(k) \setminus U_x(k - 1)$, with $S_x(0) := \{x\}$. □

***Definition 2.*** *Node-shell dimension.*

Let $x$ be a generic node of causet $C(N, E)$, and define $D_x(k) := \frac{\log(|S_x(k)|)}{\log(k)} + 1$. We say that $C(N, E)$ has *node-shell dimension* $D_x := \lim_{k \to \infty} D_x(k)$, starting from node $x$, when such limit exists, and that the causet has *node-shell dimension* $D$ whenever $D_x = D$ for all $x \in N$. □

In Figure 4 we estimate the node-shell dimension of the three representative causets shown on the left column, starting from node 1, the root.

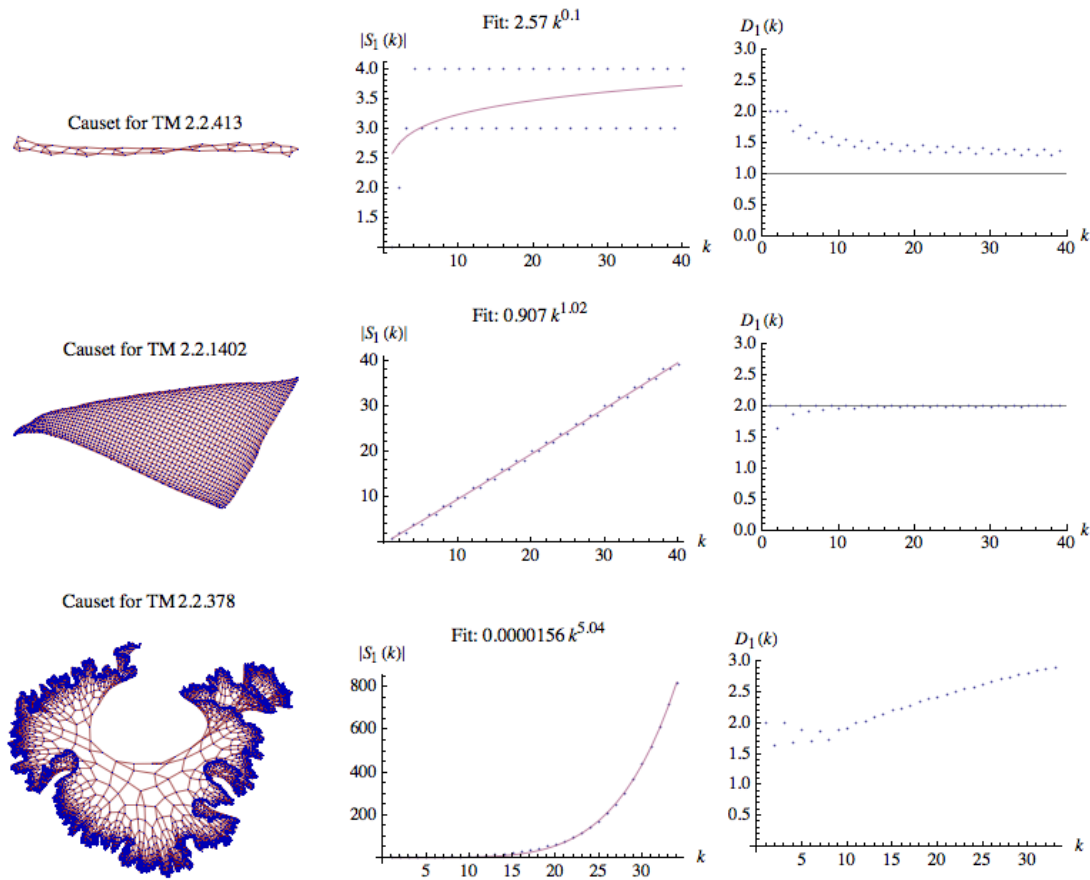

Figure 4 - Sizes $|S_1(k)|$ of node-shells at growing distance $k$ from root, with non-integer monomial fit (center), and $D_1(k)$ values (right), for three causets (left) from elementary Turing machines.

The actual node-shell sizes at progressive distance *k* are plotted in the central column, jointly with a two-parameter best fit approximation by function $f(k) = a * k^b$. The latter is a *non-integer,* monomial approximation, in which parameters *a* and *b* are real numbers. The column on the right shows the values of $D_1(k)$ at progressive distance *k* from node 1, the root. Note that we shall always trim these data sets for removing the oscillations occurring when reaching the growth boundary. Inspection of the first two causets in figure immediately reveals their respective one-dimensional and two-dimensional nature, making the associated plots somehow unnecessary. But we still present them for illustrating a point about their usage for obtaining our dimensional estimates. When $|S_x(k)| = a * k^b$, then $D_x := lim_{k \to \infty} (\frac{\log(|S_x(k)|)}{\log(k)}+1) = b + 1 + lim_{k \to \infty} \log_k a = b + 1$: thus, the exponent of a satisfactory non-integer monomial approximation would provide, in the limit of $k \to \infty$, an alternative way to estimate the node-shell dimension. However, the $\log_k a$ term in the above equation is far from negligible, considering realistic values of distance *k*, and relative to the dimensional values that we shall be dealing with -- typically small integers. The situation is illustrated in Figure 5, which plots the value of $\log_k a$ as a function of *k* with parameter *a* fixed: this value represents the discrepancy between the estimate (*b*+1) derived from the exponent of the monomial approximation, and the estimate $D_1(k)$ derived from Definition 2.

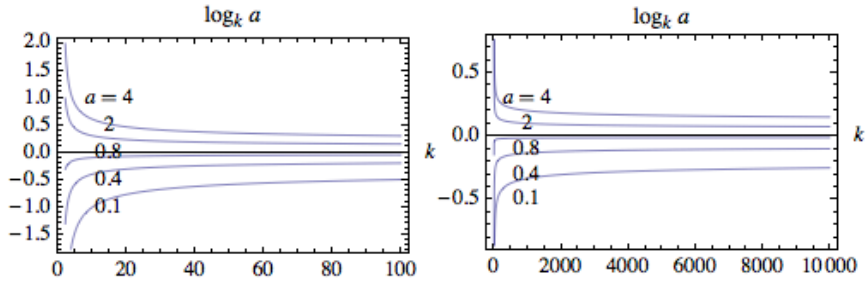

Figure 5 - The $\log_k a$ discrepancy between two alternative dimensional estimates, as a function of distance *k* and parameter *a*, assuming a non-integer monomial growth $a * k^b$ of node shell sizes $|S_1(k)|$. Gap shown for distances up to k=100 (left) and k=10,000 (right).

When the non-integer monomial approximation $a * k^b$ happens to yield a value $a ≈ 1$, then the difference between the two estimates is indeed negligible, as it happens with the example in the second row in Figure 4; in this case, $a = 0.907$, and dimension 2 is revealed both by exponent $b = 1.02$ and by the rapidly converging values of $D_1(k)$. However, when $a ≉ 1$ the slow convergence of $D_1(k)$ becomes effective, and the estimate based on exponent *b* becomes more useful, as it happens with the example in the first row of Figure 4, and in some less trivial cases found later. The diagrams in Figure 5 also show that considering much larger (and impractical) distances would provide limited benefit.

In the third case of Figure 4, for TM 378, the node-shell dimension estimator fails: $D_1(k)$ diverges, and the monomial approximation appears to succeed, but only through a vanishing value of multiplicative parameter *a,* and an ever growing value of exponent *b*, as the size of the data set increases. These facs are indications of the predominantly exponential nature of function $|S_1(k)|$, as confirmed by applying a two-parameter exponential fitting function $f(k) = c * d^k$, as shown in Figure 6.

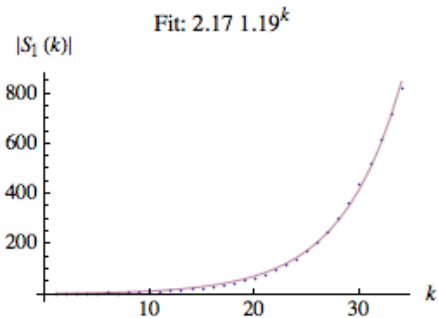

Figure 6 - Exponential fit for the sizes $|S_1(k)|$ of node-shells at growing distance *k* from the root, for the causet of TM 378.

In summary, our empirical approach to determining the node-shell dimension of our causets is as follows. First we discriminate between polynomial and exponential growth, by using the two-parameter fitting functions $a * k^b$ (non-integer monomial) and $c * d^k$. Then:

- We recognize a *polynomial growth* whenever the *a* parameter assumes rather stable, non-vanishing values. If this is the case, and the oscillations of $|S_x(k)|$ with *k* are small, we assign to the causet a (finite) dimension *b*+1, based on the exponent of the fit. The $D_x(k)$ plot may still be useful for ruling out an exponential growth, since it will not show signs of convergence. However, these diagrams can be quite misleading when trying to infer a reasonably accurate value of their finite limit (see, e.g., the upper-right diagram in Figure 4), due to the above discussed $\log_k a$ term. Thus, we shall mainly look at the *b* parameter.

- We recognize an *exponential growth* whenever the parameters *a* and *b* of the monomial fit tend, respectively, to vanish and to steadily grow, as the distance *k* increases, while the exponential fit parameters *c* and *d* assume relatively stable values of the order of magnitude of 1; correspondingly, the $D_x(k)$ plot shows a steady, roughly linear growth. In this case, Definition 2.2 is not useful, since it would assign a meaningless, *infinite* value to the causet dimension, and we are forced to *suspend our judgement about dimension.* (Other dimension estimators would be needed, and some are indeed available [11, 18], but we shall nod discuss them here.)

While an exponential node-shell growth does not seem to help with dimension estimation, we can still use it as an indicator of a *negative curvature* of the causet. Let us take a closer look at the issue of graph curvature, by considering a well known definition of curvature for planar graphs.

***Definition 2.3. Combinatorial curvature for planar graphs.***

For a planar graph $G(N, E)$, the *combinatorial curvature* of a node $x \in N$ is defined as

$$\text{cc}(x) := 1 - \frac{\text{degree}(x)}{2} + \sum_{f \sim x} \frac{1}{\text{size}(f)}$$

where summation is over all faces $f$ incident with $x$. □

Let us apply the above definition to the causet for TM 378. Note that this causet exhibits a highly symmetric structure, whenever the number of nodes is $N_h$, corresponding to an $(N_h - 1)$-step computation, where $N_0 = 1$ and $N_i = 2 N_{i-1} + 2i - 1$, for $i > 0$. An example ($h = 5$) is shown in Figure 7, where an alternative layout for the causet of a 114-step computation is provided. All edges are oriented left to right. Index $h$ corresponds to the 'height' of the central vertex.

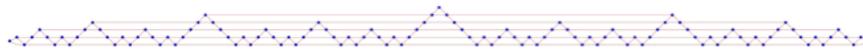

Figure 7 - Alternative layout for the causet of a 114-step computation of TM 2.2.378.

Let us take a closer look at the issue of graph curvature, by considering a well known definition of curvature for planar graphs.

Consider now the family of completely symmetric graphs for TM 378, of the type illustrated in Figure 7; in these causets one finds faces of sizes 3, 4, 5, and one external face of variable size. Nodes exhibit 12 different values of combinatorial curvature, both positive and negative; their distribution is shown in Figure 8 for the case $h = 10$ (the overall shape is basically unaffcted by changing $h$).

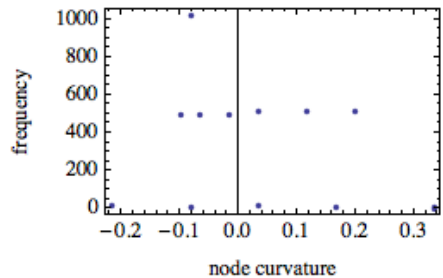

Figure 8 - Distribution of node curvature in a symmetric causet for TM 378, with $N_{10} = 4073$ nodes.

Note that the average curvature is *always positive*, and tends to zero as the graph size grows. This is an immediate consequence of the fact that, for a planar graph $G(N, E)$, letting $F$ be the set of faces, we have

$$\sum_{x \in N} \text{cc}(x) = |N| - |E| + |F| = 2.$$

The first equality is established by simple counting, and the second one is Euler's formula.

Is there a contradiction between attributing *negative* curvature to a graph with exponentially growing node-shell sizes, while detecting a *positive*, average *combinatorial* node curvature in it (tending to 0)? No. The situation is best explained by considering the regular tiling of the hyperbolic plane with Schlafli code {5, 4}: tiles are regular pentagons with right angles, and 4 tiles meet at each vertex. Figure 9 (taken from Don Hatch's web site http://www.plunk.org/~hatch/HyperbolicTesselations/) shows this tiling on the Poincare' disc, which is a model of the hyperbolic plane.

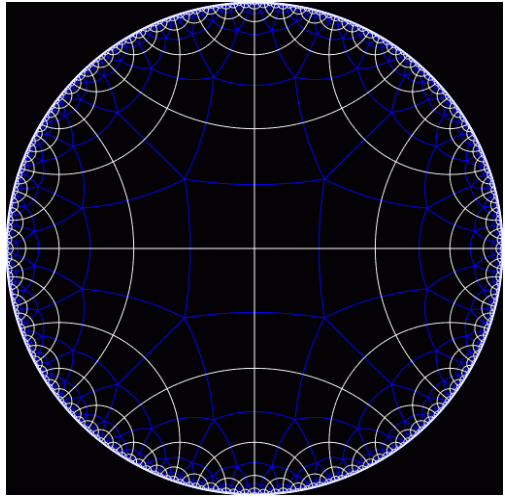

Figure 9 - Hyperbolic planar tessellation with Schlafl code {5, 4} (white graph) and its dual, with code {4, 5} (grey graph). Figure from http://www.plunk.org/~hatch/HyperbolicTesselations/

Margenstern and Morita [8] show that the pentagons in a quadrant of this tessellation can be arranged, according to their bordering relation, into a so called *Fibonacci tree*. The tree, which is indeed a graph, is represented by grey lines in Figure 9, and defines the dual tessellation {4, 5}; its name is justified by the fact that, considering one quadrant of the disc, the number of nodes at level *n*, the root being at level 1, is $N(n)$ = Fibonacci(2*n*). Thus, we can conclude that the node-shell growth for this graph is exponential, with base 1 + GoldenRatio = 2.618.

In this infinite graph with code {4, 5} all nodes are equivalent and have negative curvature -1/4. But if we consider a *finite* portion of this planar graph, of some fixed radius, then the 'leaves' have positive curvature, whose precise value depends on the actual size of the external face, and completely balance the negative curvature of the internal nodes. Note that in hyperbolic tessellations with constant curvature there is not a one-to-one correspondence between the latter and the basis of the exponential node-shell growth: for example, tessellation {3, 7} also exhibits a 1 + GoldenRatio node-shell growth, but local curvature is -1/6.

The causet of Figure 7 is not as uniform as the above tessellations; nodes with several different values of positive and negative curvature are found both on the border and internally, but the study of node-shell growth starting from different nodes, even considering undirected arcs, cannot but confirm the observed exponential behavior. (In the sequel we shall mainly concentrate on node-shell dimension starting from the root.)

## 2.2 Causets for more complex Turing machines

By adding just one element (or *color*) to the tape alphabet, and considering 2-state, 3-color Turing machines -- a family of $(2*2*3)^{(2*3)}$ = 2,985,984 elements -- causet complexity increases. The vast majority of these graphs repeat the three fundamental patterns illustrated in Figures 2, 3 and 4, with minor variants, as shown in Figure 10.

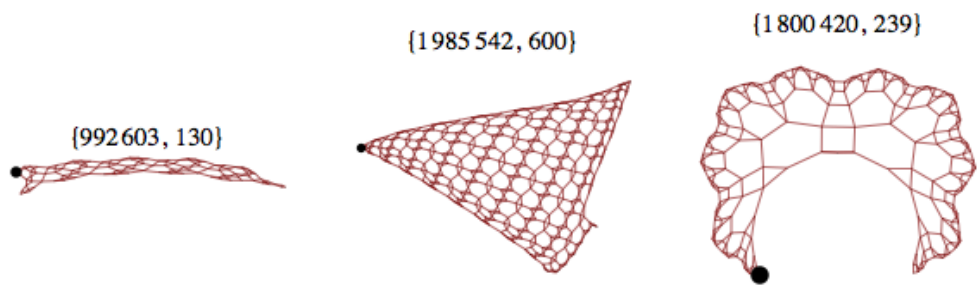

Figure 10 - Causets for three 2-state 3-color Turing machines. Labels indicate the code number of the machine and the number of computation steps.

An interesting exception is the causet for TM 2.3.621900, shown in Figure 11.

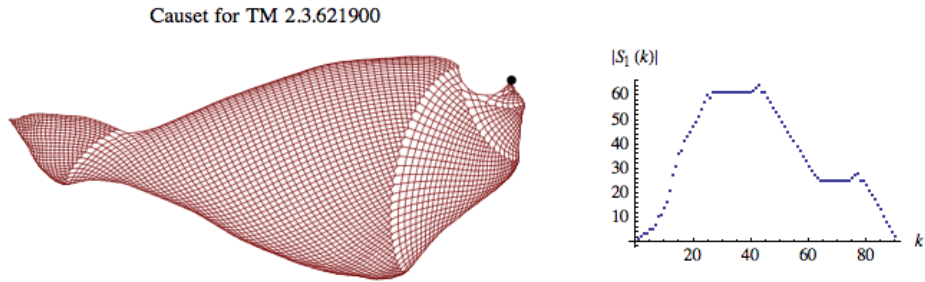

Figure 11 - The causet for 2-state, 3-color Turing machine 621900 formed by a progression of triangular 2D grids of *exponentially* growing size (left), and its node-shell growth (right).

This causet is organized as a sequence of triangular *layers,* consisting of triangular, 2D grids of exponentially growing size; four of them are clearly visible, and a portion of the fifth appears on the left, which will grow upwards and to the right. An interesting feature of this graph is that, starting from the root and taking the path at the upper-right border of the planar layout, one can reach a new layer every second step. As a consequence, node-shell growth appears in this case similar to the one observed in a 3D cubic lattice restricted to positive $x$ coordinates, in which every step along the $x$ semi-axis gives access to a new 2D subspace -- a square, $yz$-grid.

A similar causet is obtained from a 3-state, 3-color Turing machine, namely TM 3.3.61786677050, as shown in Figure 12; a 'fast' path is still avaliable (the lower border in the layout), that accesses a new triangular segment every second step, but in this case layer sizes grow linearly.

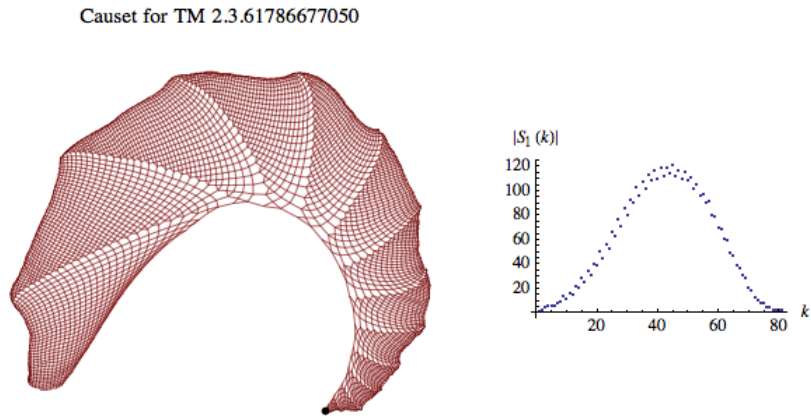

Figure 12 - The causet for 3-state, 3-color Turing machine n. 61786677050 formed by a progression of triangular 2D grids of *linearly* growing size (left), and its node-shell growth (right).

Interestingly, in spite of the different *layer* growth rate -- exponential vs. linear -- both of these causets exhibit node-shell dimension 3, starting from the root, and appear as a sort of planar, *two-dimensional implementation of three-dimensional space*. The diagrams at the right of Figures 11 and 12 do not completely reflect this result, since the 'fast' path at the border of the causet quickly exposes the incompleteness of these finite approximations to infinite graphs, but the approximately quadratic growth for a prefix of the sequence $|S_1(k)|$ is apparent in the plots of Figure 13.

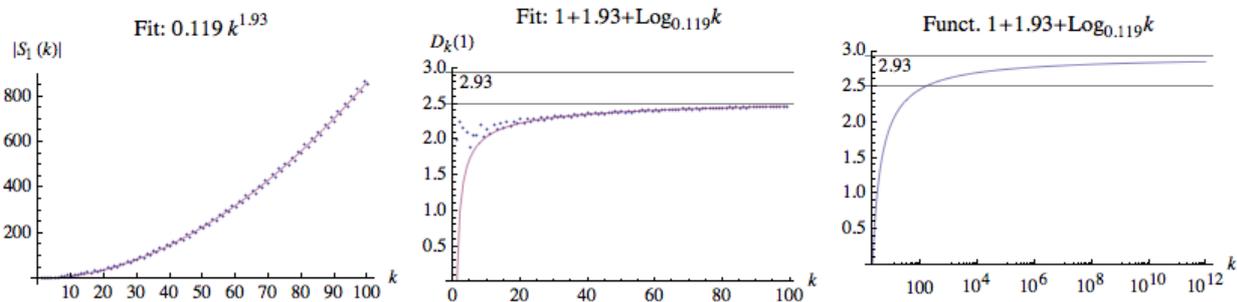

Figure 13 - Sizes $|S_k(1)|$ of node-shells at growing distance $k$ from root, with non-integer monomial fit (left). $D_1(k)$ values, compared with function $1 + 1.93 + \text{Log}_{0.119} k$ (center). The same function is shown in a LogLinear plot, showing correct convergence to 2.93 (right). Data for the causet from TM 3.3.61786677050.

These plots illustrate well the previously discussed mismatch between the two alternative dimensional estimates, related to the $\log_k a$ term. The monomial fit technique (left) detects dimension $1.93 + 1 = 2.93$. The plot in the center may suggest convergence to 2.5, but, assuming $a = 0.119$ and $b = 1.93$, function $1 + b + \text{Log}_a k$, which matches the central plot up to $k = 100$, indeed converges to the expcted value 2.93, as shown by the LogLinear plot on the right.

The fact that node-shell dimension is basically unaffected by the choice between a linear and an exponential *layer* growth rate is confirmed by analyzing simplified versions of the two layered graphs under consideration. Let $D(i)$ be the *disk* of radius $i$, that is, a subgraph of the undirected 2-D square grid in which every node has at most Manhattan distance $i$ from the central node, and let $G_{\text{lin}}$ (resp. $G_{\text{exp}}$) denote an infinite, undirected graph obtained by arranging a sequence of such disks $D(i)$ with $i$ growing linearly (resp. exponentially), so that only their central nodes are connected, as shown in Figure 14.

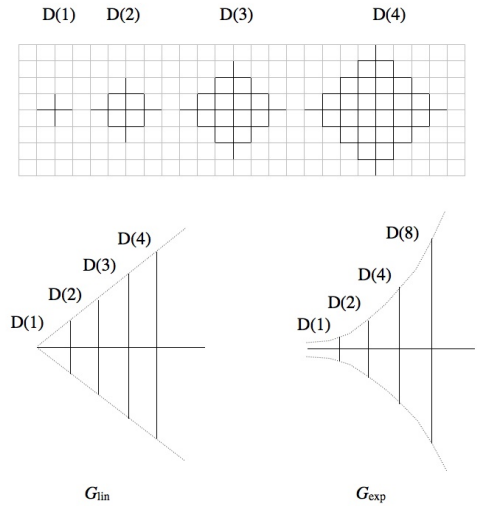

Figure 14 - Graphs $G_{\text{lin}}$ and $G_{\text{exp}}$ built by interconnecting disks with radii growing, respectively, linearly and exponentially

For graph $G_{\text{lin}}$, node-shell sizes at *even* distance $2d$ from the root (taken as the center of the smallest disk) are precisely $|S_{2*d}(1)| = 1 + \sum_{i=1}^{d} 4i$ ($d = 0, 1, ...$). Thus, node-shell growth is $O(d^2)$, and the graph dimension from the root is 3. Analogously, for graph $G_{\text{exp}}$, assuming disk radius growth function $r(d) = \lfloor b^d \rfloor$, with $b>1$, node shell sizes at distance $d$ are approximately $|S_d(1)| = \sum_{i=1}^{h(d)} 4i$, where $h(d)$ is the ordinate of the intersection point between linear function $y_d(x) = d-x$ and exponential function $r(x) = b^x$. By solving, we obtain:

$$h(d) = \frac{\text{ProductLog}\left[b^d \, \text{Log}[b]\right]}{\text{Log}[b]}$$

(Function *ProductLog*[$z$] is also known as the Lambert *W* function.) But function $h(d)$ is asymptotically linear, thus, again, node-shell growth is $O(d^2)$, and the graph dimension from the root is 3.

The family of 2-state, 3-color Turing machines includes a remarkable member. TM 2.3.596440, also known as Wolfram 2, 3 TM, is the smallest known universal Turing Machine (although the proof, due to Alex Smith [20], is based on a controversial notion of universality). Perhaps not surprisingly, the causet for this machine has a very peculiar structure, as shown in Figure 15 (left).

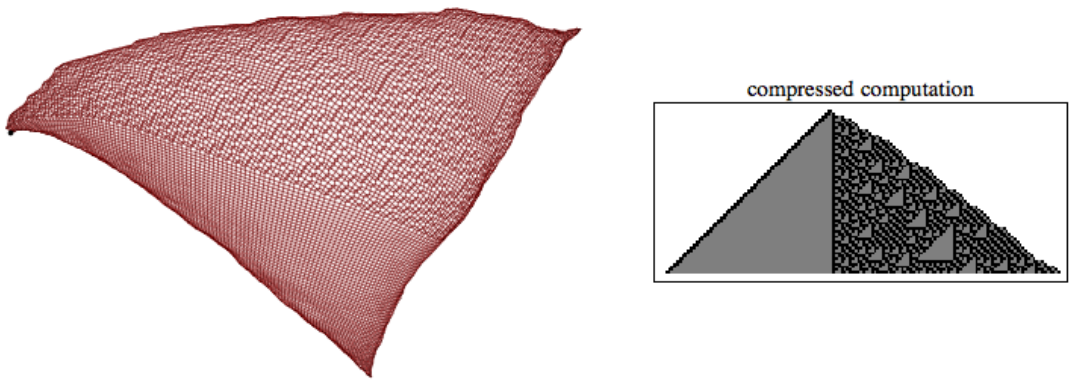

Figure 15 - The causet for universal TM 2.3.596440, and the compressed computation

In [21] a *compression technique* is adopted for obtaining compact respresentations of computations from TMs and other models, and for possibly exposing their pseudorandom character. For TMs this technique typically consists in stacking only those tape configurations at which the control head is further to the left (or right) than ever before. For the specific machine under consideration, the same irregular pattern of triangles emerges both in the causet and in the compressed computation, as shown in Figure 15. This is a remarkable, but fortuitous coincidence. It is remarkable because the two structures are built rather differently: one describes the causal structure of events, without explicit representation of tape content, and the other is formed by a subsequence of tape configurations. And it is fortuitous because, for example, compressing the computation of elementary TM 378 provides a poor, linear representation of the computation which is completely different from the corresponding, hyperbolic causet shown in Figure 4. Only the latter seems to adequately reflect, with its curvature, the nested structure of the computation. Furthemore, the computations of TM 378 and TM 1402 basically yield the same compressed pattern, thus revealing the weak discriminating power of the compression technique.

In conclusion, we do not exclude that alternative, carefully chosen compression criteria could prove more effective in characterizing computations, but they would be ad-hoc; on the contrary, causets offer a technique of general applicability, and involve no arbitrariness, for capturing the essential features of computations and effectively discriminating between them: causets appear as ideal candidates for representing universal computations and letting them play some role in fundamental physics.

Other interesting examples of TMs have been identified in [21], based on the compression technique. In Figure 16 we present the causets that we have derived from some of them.

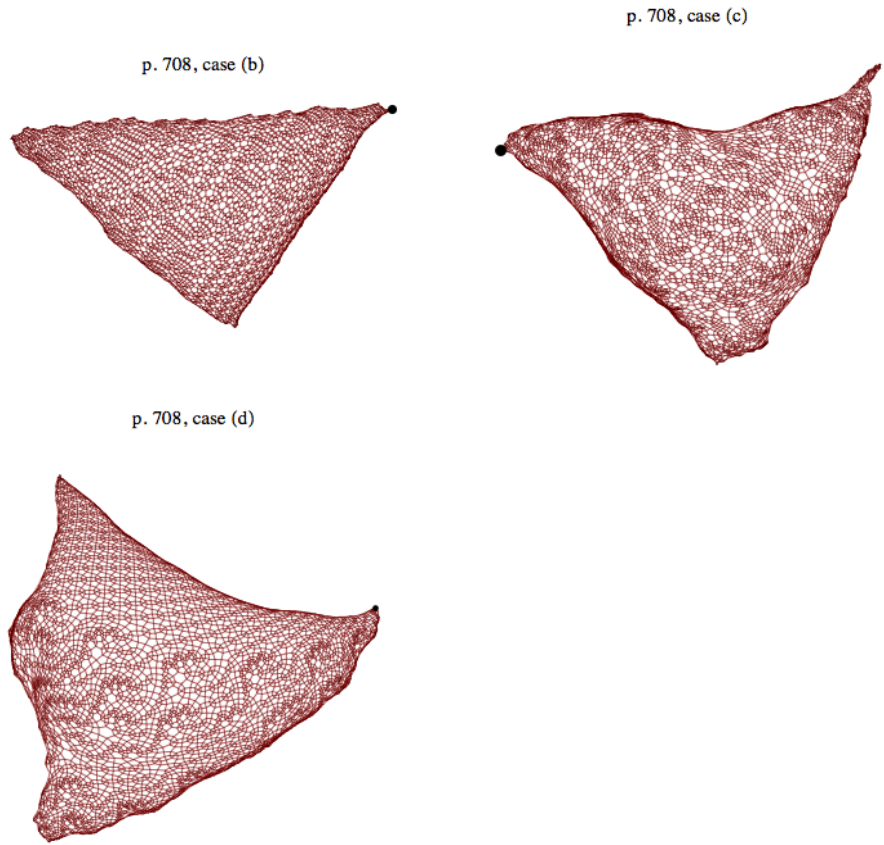

Figure 16 - Causets for three TMs from [21], p. 708.

Dimension and curvature are *quantitative* properties, but their analysis does not appear as the only key to investigate the emergent properties of the causets from Figures 15 and 16, in all of which node-shell growth is roughly linear. The properties that become relevant here are, rather, of *qualitative* nature: what makes them special is the peculiar mix of regularity and pseudorandomness, and the partition into a smaller periodic component and a larger pseudo-random one, which is also exhibited by the well known elementary cellular automaton (ECA) 'Rule 30' shown in Figure 17.

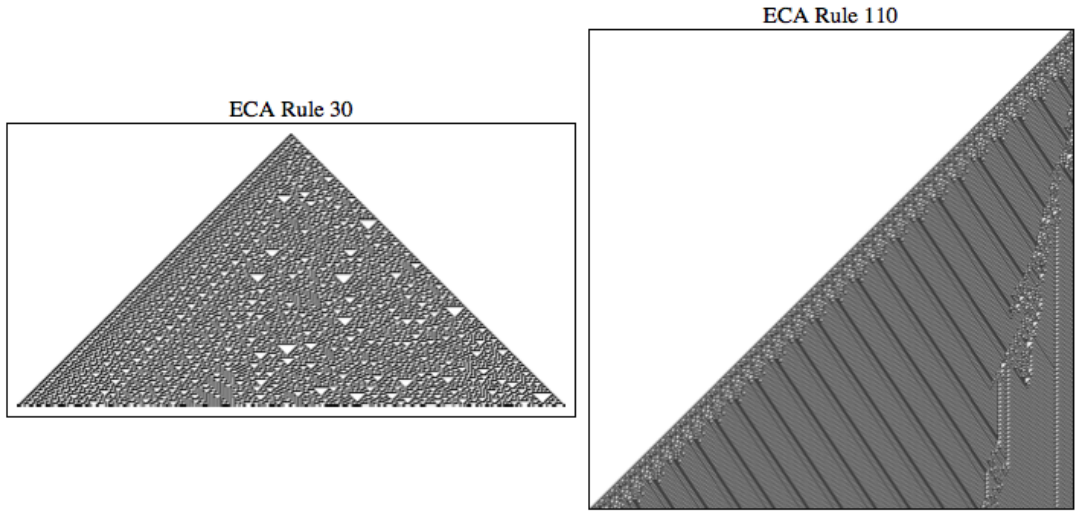

Figure 17 - Elementary cellular automata: Rule 30, with a smaller periodic and larger pseudo-random sections, and Rule 110, with particles.

The comparison between causets and cellular automata is indeed quite appropriate, since also the latter can be seen as spacetime structures, with associated notions of causality, maximum information transmission speed, light cones, and so on. For example, the typical pattern of scattered triangles observed in some one-dimensional cellular automata, where triangles may appear uniform or themselves filled by a regular pattern, also appears in two of the above causets. But one of the most remarkable features of cellular automata (in 'class 4', according to Wolfram's classification) is the emergence of particles, or 'gliders', intended as periodic, spatially confined structures that move and interact in complex ways, on top of a periodic background, as it happens with ECA Rule 110 (Figure 17, r.h.s.). Thus, it would seem reasonable to look for the presence of similar, localized periodic structures in causets too, and some of the graphs in Figure 16 apparently lend themselves to this search. We shall discover more evident cases of causet particle later, when dealing with computations on higher-dimensional supports.

# 3. Causets from computations on a linear support

Several simple models, beside Turing machines, operate on a linear support. In this section we define causets for the computations of four of them -- mobile automata on tape, string rewrite systems, tag systems and cyclic tag systems, and show that all of them are planar.

## 3.1 Mobile automata on tape

In mobile automata (MA), originally proposed in [21] as an intermediate model between elementary cellular automata and Turing machines, a *stateless* control head positioned at tape cell $c$ rewrites it and moves to the left or right depending on the current content of $c$ and of its two neighbors. An example of a mobile automaton rule and a corresponding computation, borrowed from [21], is provided in Figure 18. We conveniently call this model 3x1 MA (for 3-read, 1-write).

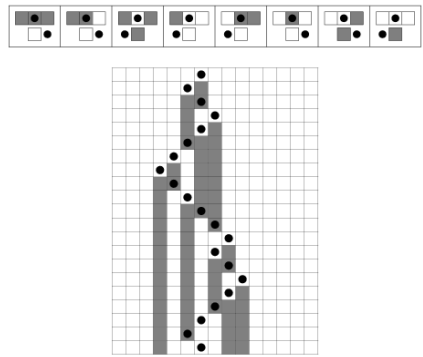

Figure 18 - A Mobile Automaton rule (upper) and a computation segment (lower), showing stacked tape configurations (rows) and current control head positions (black circles). (Figure from [21], p. 71.)

**Causet construction** (see Figure 19).
- Every step of the MA becomes an event of the causet.

- A directed edge connects event *e1* to event *e2* iff *e2* reads a cell that was written by *e1* and by no other event in between.

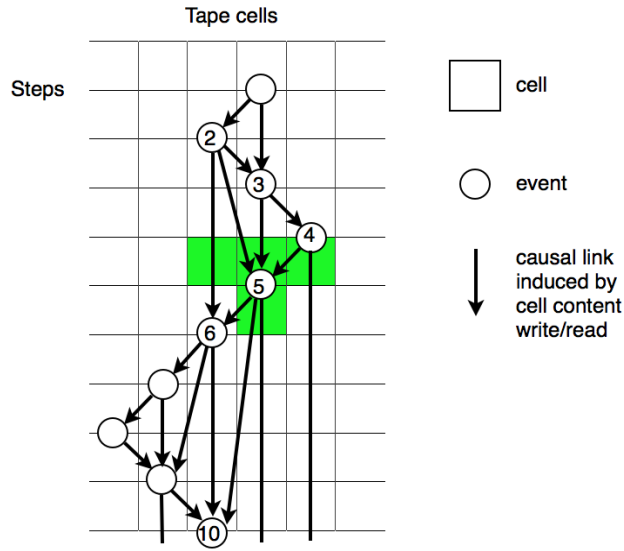

Figure 19 - Causet construction for a 3x1 mobile automaton on tape.

Figure 19 refers to the first 10 steps of the computation in Figure 18. Note that the above construction implies that every causet node, corresponding to an event *e* that reads cells $c_{i-1}$, $c_i$, $c_{i+1}$ and writes cell $c_i$, has at most three incoming arcs -- as many as the most recent events that have written some or all of the three cells; for example, event 5 reads the three shaded cells that have been written most recently by events 2, 3 and 4. On the other hand, there can be an arbitrary number of outgoing arcs, since cell $c_i$ can be read, *without being written*, by an arbitrary number of events, before being read *and written* again by a further event. For example, event 5 writes a shaded cell that is subsequently read by events 6 and 10: in principle, nothing would prevent further events to read it, before the occurrence of a rewrite event for that cell.

**Proposition 2.** *The causet of a 3x1 mobile automaton on tape is planar.*

Proof. Consider the specific layout in which all causet arcs are straight line segments, and nodes are arranged in columns associated with tape cells, as in Figure 19. By definition, arcs can only join nodes in the same column (vertical) or in two adjacent columns (oblique). The absence of arc crossings in this layout is proved by contradiction. Assume arcs $a \to b$ and $c \to d$ cross each other. Clearly both of them must be oblique. Assuming, w.l.o.g., that *a* is the first event to occur, and $a \to b$ moves left, there are three possible arrangements of arc $c \to d$ w.r.t. $a \to b$, as shown in Figure 20. All three cases violate the causet construction rule by which a directed edge connects event *e1* to event *e2* iff *e2* reads a cell *x* that was written by *e1* and by no other event in between: in the first case, event *c* rewrites cell *x* that was written by *a*, and is therefore incompatible with arc $a \to b$; in the second case, it is event *d* to be incompatible with arc $a \to b$; in the third case, arc $c \to d$ is incompatible with event *b*. □

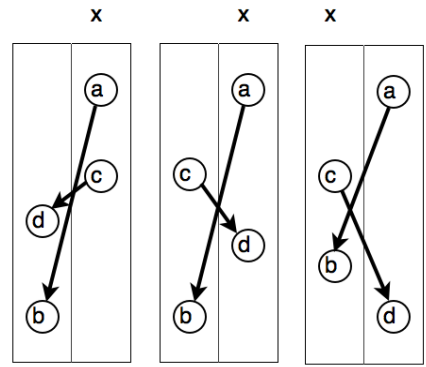

Figure 20 - Impossible arc crossings in the straight line layout of causets from 3x1 mobile automata on tape.

The planarity result above is valid also for *jumping* mobile automata, in which the control point is allowed to move on the tape by steps longer than a unit, provided that the automaton rule be still based on reading cell triplets, since the latter is a sufficient condition for having arcs that span at most 2 columns. A difference between the *jumping* and the original ('*walking*') model is that causets from the latter are always *totally* ordered (as with Turing machines), since the cell triplet read by one event always includes a cell written by the previous event, while causets from jumping machines are in general partially ordered in strict sense.

A generalization of the MA model is considered in [21] that allows each step to rewrite the *complete* triplet of cells that has just been read. We call them 3x3 MA. The planarity result easily extends to this generalization, as established below.

**Proposition 3.** *The causet of a 3x3 mobile automaton on tape is planar.*

Proof. A planar layout can be immediately obtained by considering the usual stack of tape configurations and by identifying each event with the triple of cells that are read/written at each step, as shown in Figure 21. Then, causet arcs can only be vertical and non-overlapping. □

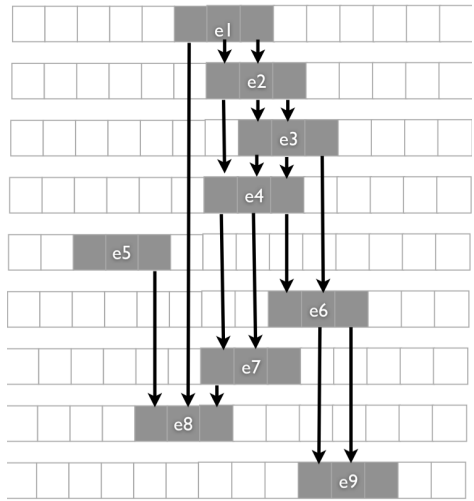

Figure 21 - Causet construction for a 3x3 mobile automaton on tape.

Again, the above result applies to walking as well as jumping automata, with the causets from the former being in fact totally ordered.

The family of 3x1 walking MA on binary tape has 65,536 elements and has been exhaustively explored in [21]: all of these exhibit simple, regular behaviour, while no case of pseudo-randomness is found. Causets are correspondingly regular, as shown in Figure 22, where MA jumps have length at most 3. The only novelty w.r.t. causets from Turing machines is the presence of nodes with higher, possibly unbounded degree (central graph in figure). Note that, by allowing sufficiently long jumps, causets may become disconnected, as it happens in the third example; this might as well occur with jumping Turing machines.

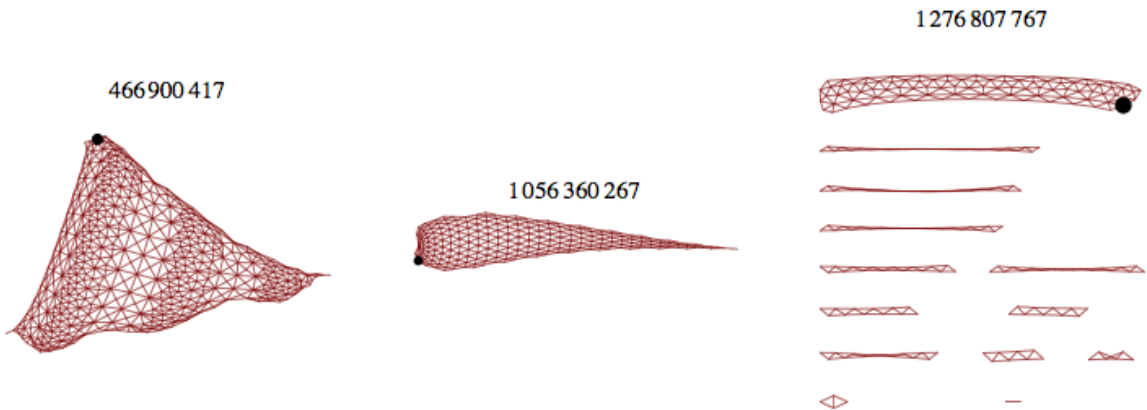

Figure 22 - Three causets from 3x1 MA computations

Wolfram [21] has also investigated the family of 3x3 walking MA on binary tape, which includes over 4 billion elements, and has found that the vast majority of them yield regular, repetitive or nested behavior. However, two cases of pseudorandom behavior are also found, and we provide their causets in Figure 23.

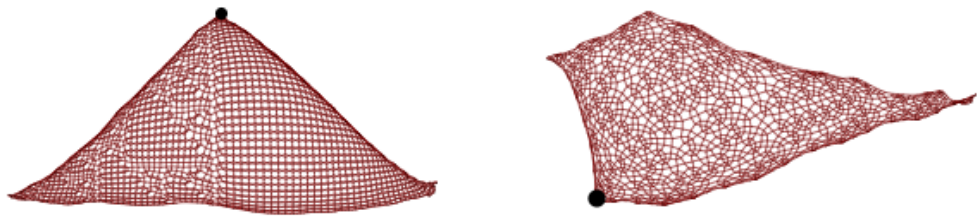

Figure 23 - Causets for the 3x3 MA from [21], p. 74 (left) and p. 75 (right).

The strong similarity between these causets and some of those derived earlier from TM computations can be taken as an indication of the provably equivalent computational power of the two models.

## 3.2 String rewrite systems

In a string rewrite system, a string on a given alphabet is rewritten according to an ordered list of rewrite rules. Rule application can follow two alternative criteria.

With *parallel* rule application, one starts scanning the string from left to right, looking for the first substring $h$ that can be matched by the *head h* of some rewrite rule $h \to b$; once the substring is replaced by the rule *body b*, the scan proceeds in the same way with the remainder of the string.

With *sequential* rule application, once the first replacement is performed, the scan terminates, and a new scan starts from the beginning of the string.

When some rule $h \to b$ is applied, we say that substring $h$ is read and substring $b$ is written.

The causet construction is valid both for parallel and for sequential rewrite systems, and is essentially illustrated in [21, p. 516], from where we borrow Figure 24.

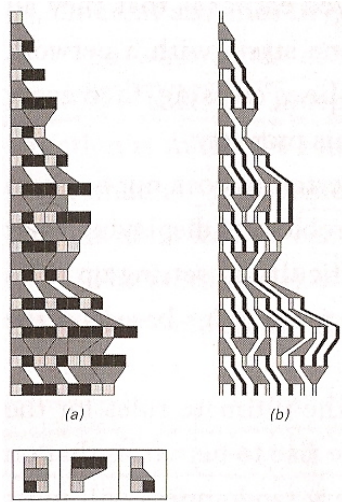

Figure 24 - Causet construction for (parallel) string rewrite systems (from [21], p. 516).

**Causet construction**

- Every rewrite rule application at every scan becomes an event of the causet.
- A directed edge connects event $e1$ to event $e2$ iff $e2$ reads a string symbol that was written by $e1$.

*Proposition 4.* *The causet of a string rewrite system is planar.*

Proof. Trivial. Diagram (b) in Figure 24 is a representation of the causet, where nodes are drawn as grey polygons and arcs as black or white thick lines. The graph is planar by construction. □

Both Turing machines and mobile automata are based on the moves of a control unit on the rigid, linear structure of a tape, and on reading a *fixed* number of cells. String rewrite systems are considerably different, for the absence of a control unit and the fact that rewrite rules for a given system usually involve heads and bodies of a variety of different lengths. A systematic exploration of this model is harder, but we can take advantage, again, of the selections carried out in [21]. What we find is that, in spite of the novel aspects of this model, the same fundamental regular causet structures -- 1D, 2D flat and hyperbolic -- seem to emerge, as shown in Figure 25, in which we consider three rewrite systems on binary alphabet involving, respectively, one, two and three rules, with specific initial conditions, as indicated in the graph labels.

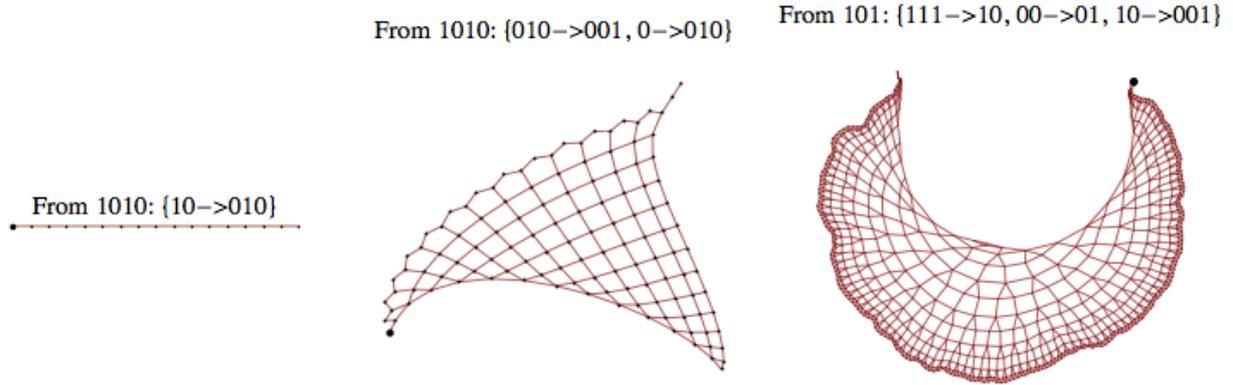

Figure 25 - Three causets for some string rewrite systems from [21], pp. 89-92.

We also find examples of causets structured as a sequence of growing 'layers'. One of them is shown in Figure 26. It is analogous to the causet from Turing machine TM 3.3.61786677050 (see Figure 12), and exhibits the same node shell growth rate, yielding the same dimension 3. Again we find that the monomial fit is a convenient technique for determining the dimension, compared with the misleading plot of the $D_1(k)$ values (compare with Figure 13).

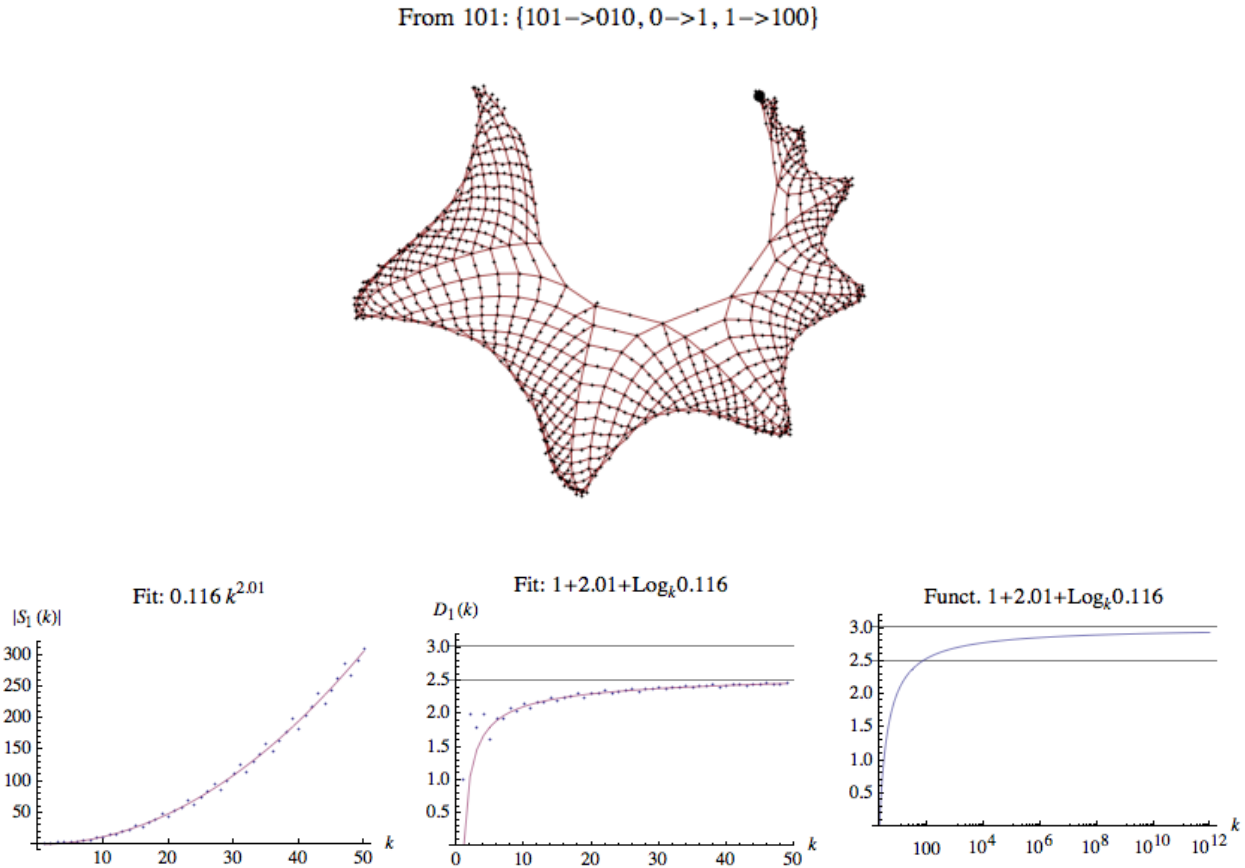

Figure 26 - The upper graph is the layered causet of a string rewrite system from [21, p. 91, case (c)], running for 1000 steps (compare with Figure 12). The plots below it refer to a 20,000-step computation, and show: the sizes $|S_1(k)|$ of node-shells at growing distance *k* from the root, with non-integer monomial fit (left); $D_1(k)$ values, compared with function $1 + 2.01 + \text{Log}_k 0.116$ (center); the same function in a LogLinear plot, showing convergence to 3.01 (right).

Out of the 10 string rewrite systems identified in [21, pp. 89-92], we single out two that exhibit interesting *qualitative* properties. They are shown in Figure 27.

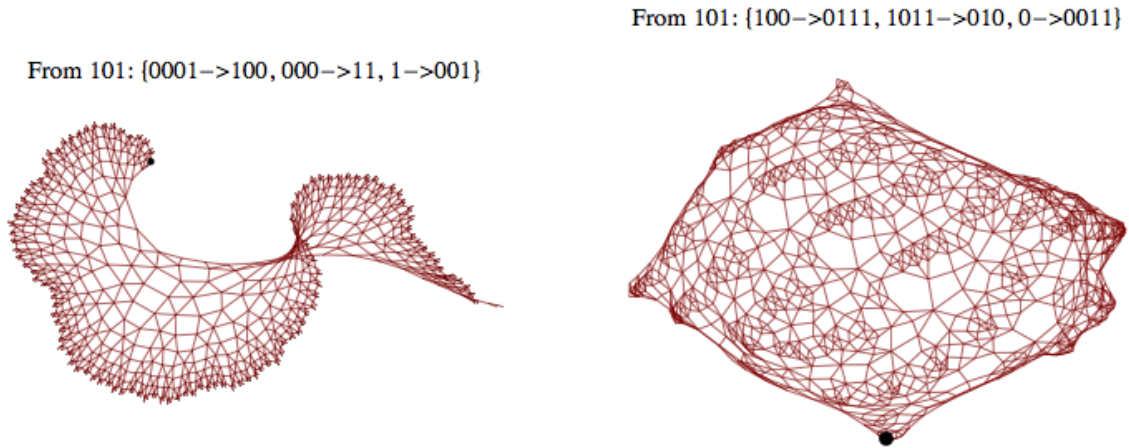

Figure 27 - Causets from 1000-step computations of string rewrite systems (g) and (h) from [21, p. 91].

The causet on the left has an overall hyperbolic structure, but does not show an obviously regular pattern as the previous hyperbolic cases; additional evidence of its non-regular character can be obtained in various ways, e.g. by considering the progression of node-shell sizes and then their differences, or higher degree differences (repeated differences of differences).

The causet on the right exhibits a complex texture at the short scale, although on a larger scale it ends up revealing a regular, nested, spiraling structure.

## 3.3 Tag systems

In a tag system, a string on a given alphabet is iteratively rewritten according to a set of rewrite rules. A fixed number $k$ of elements is dropped from the front of the string, that matches the head of some rule $h \rightarrow b$, and then the rule body $b$ is appended ('tagged') to the tail of the string. All heads have length $k$, while bodies may have different lengths.

**Causet construction**

- Every string rewriting becomes an event of the causet.
- A directed edge connects event *e1* to event *e2* iff *e2* drops (reads) a string symbol that was appended (written) by *e1*.

*Proposition 5.* *The causet of a tag system is planar.*

Proof. First we build a causet layout (see Figure 28) by stacking and shifting the successive string configurations so that symbols temporarily unaffected by the rewriting appear aligned in the same column, and by representing events as circles at the right of the string that they manipulate. Causal links are determined by the causet construction rule above. In Figure 28, for example, event 4 reads symbols *g* and *h*, that have been written by event 1, hence arc $1 \rightarrow 4$ is created (duplicate arcs can be then removed). Event 5 reads symbol *i*, also written by event 1, but it also reads symbol *j*, written by event 2, thus arcs $1 \rightarrow 5$ and $2 \rightarrow 5$ will appear. Arc overlappings can then be avoided by the layout sketched on the r.h.s. of the figure, which is possible since, with sequentially numbered events, the following holds:

$$\forall \, i : \text{Max}\{j \mid i \rightarrow j \in \text{Arcs}\} \leq \text{Min}\{j \mid (i+1) \rightarrow j \in \text{Arcs}\}. \quad \square$$

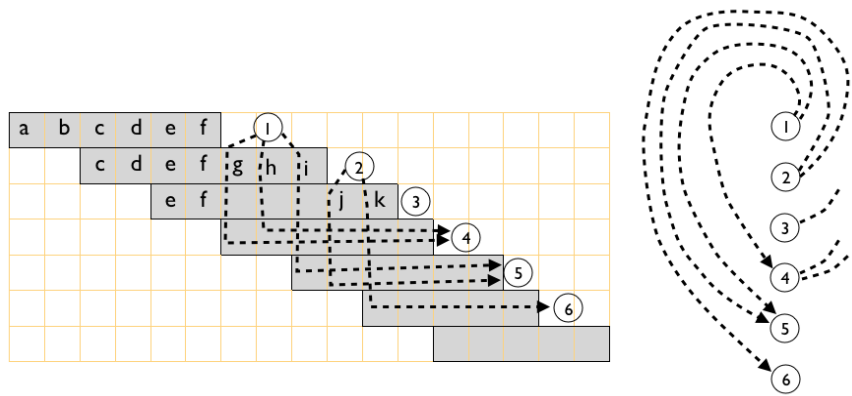

Figure 28 - Tag system computation (left) and planar embedding of causal arcs (right). Circles represent rewriting events.

When the fixed number of elements dropped from the string is $k = 1$, then every event has one incoming arc, thus causets can only be trees or forests; these provide very poor structures for a spacetime.

When $k = 2$, linear, 1-D causets are frequent, as in all previous models (Figure 29), but a novel feature emerges too, that was not observed before. While the typical movement of the Turing machine or mobile automaton control point, up and down the tape, typically yields roughly triangular planar causets, the circular rewriting policy of tag systems -- reading the *head* and writing the *tail* of the string -- leads to 'closing the triangle', and typically yields roughly circular causets, that are grown by a spiraling process. Some of these causets are shown in Figure 30, where labels indicate the initial string and the replacements for the 4 possible bit pairs; this convention for labels is valid also for some of the subsequent figures. Depending on the specific (but unessential) details of the underlying graph layout algorithm, some of these disks are rendered as conical.

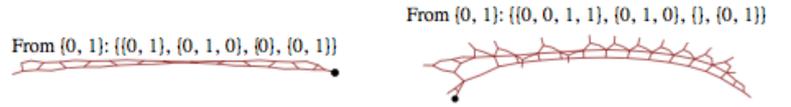

Figure 29 - Linear causets from tag systems.

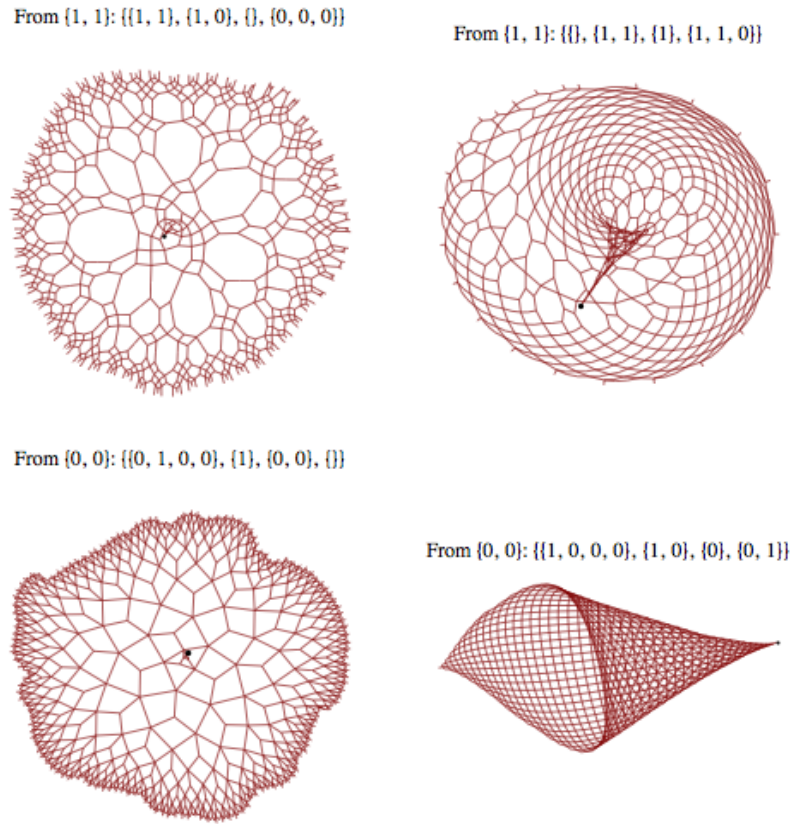

Figure 30 - Roughly circular causets from tag systems, with spiraling growth.

Node shell dimension for three circular causets of Figure 30 is investigated by the plots in Figure 31.

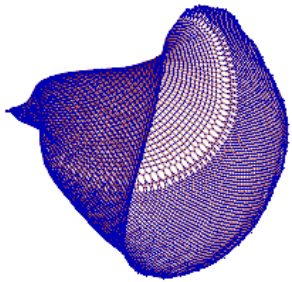 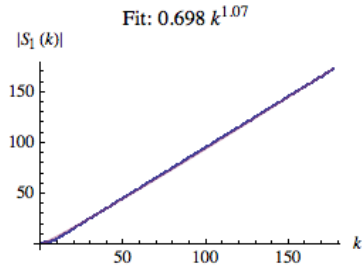 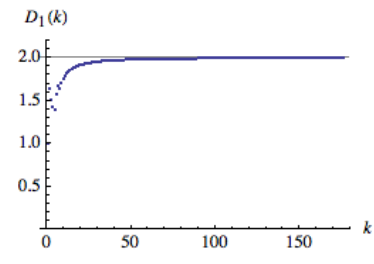

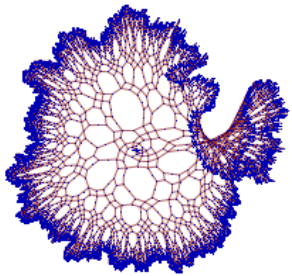 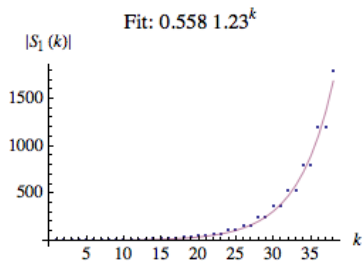 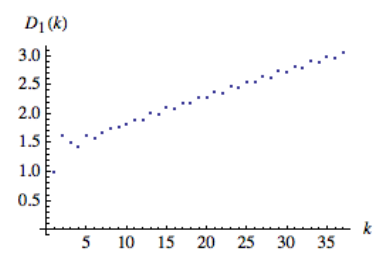

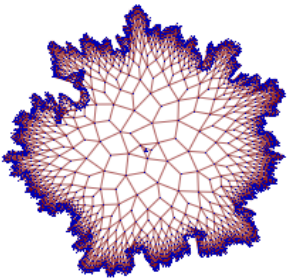 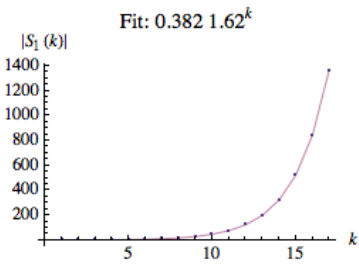 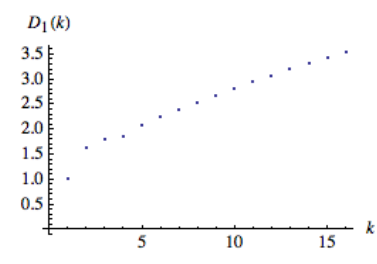

Figure 31 - Node shell growth for three circular causets from Figure 30.

The first causet has node-shell dimension 2. The second and third causets exhibit negative curvature.

The latter has remarkable properties. Node-shell sizes from the root are {1, 1, 2, 3, 4, 7, 11, 18, 29, 47, 76, 123, 199, 322, 521, 843, 1364, ...}; starting from the 4th element, this is a Fibonacci sequence, corresponding to the 1.62 basis found by the exponential best fit, a value which closely approximates the golden ratio. Even more interestingly, by ignoring the small initial transient of the first 7 events, the causet becomes uniform, in the sense that there are only 2 equivalence classes of nodes -- the degree-3 and the degree-5 nodes: all nodes in a class have an identical view of the *directed* graph, that is, exactly the same sequence of node shell sizes at growing distances. Degree-3 nodes 'see' the sequence {1, 2, 5, 9, 16, 27, 45, 74, 121, ...} and degree-5 nodes 'see' {1, 3, 6, 11, 19, 32, 53, 87, 142, ...}. Both sequences satisfy the recurrence $|S_{k+1}(x)| = |S_{k-1}(x)| + |S_k(x)| + 2$, for $k \geq 3$. This is a property also of hyperbolic planar tessellations (which define *undirected* graphs - see Figure 90), and corresponds, in the continuum, to the Riemannian notion of *space form*, which is a manifold of constant curvature.

The appearance of disk structures represents a significant departure from the previous cases. In a continuous spacetime setting, the distinction between the 'open' shape of a triangle and the disk, representing its 'closed' version, evokes a distinction between space manifolds (the spacelike components of spacetime) with and without *border*. However, we are not aware of reformulations of the notion of border from a continuous to a discrete setting, hence we leave the above remark as a pointer to further investigation.

By putting the above disk-like causets side by side with the previously found form of negatively curved causet, an interesting analogy emerges with, respectively, the Riemann-Poincaré disk and the Beltrami-Poincaré half-plane, two well known isomorphic models of two dimensional hyperbolic geometry. For the curious reader, we illustrate the growth process for these two causet forms in Figure 32, where we use an example from string rewrite systems (see Figure 25) and one from tag systems (see Figure 30). The grey levels reflect the order of event creation, module some suitably large integer. In the l.h.s. graph the growth process proceeds clockwise as a spiral; in the r.h.s. graph it sweeps the whole structure, from right to left.

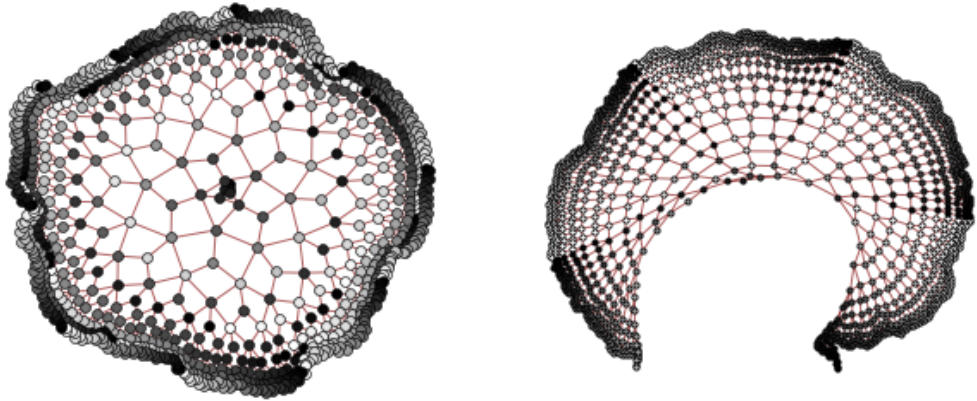

Figure 32 - Two causets with negative curvature representing finite approximations of, respectively, the Riemann-Poincaré disk model (left) and the Beltrami-Poincaré half-plane model (right) of the hyperbolic plane.

Finally, as seen with all previous models, we find highly irregular causets. One in shown in Figure 33, derived from a tag system computation singled out in [21] (p. 94, case (f)).

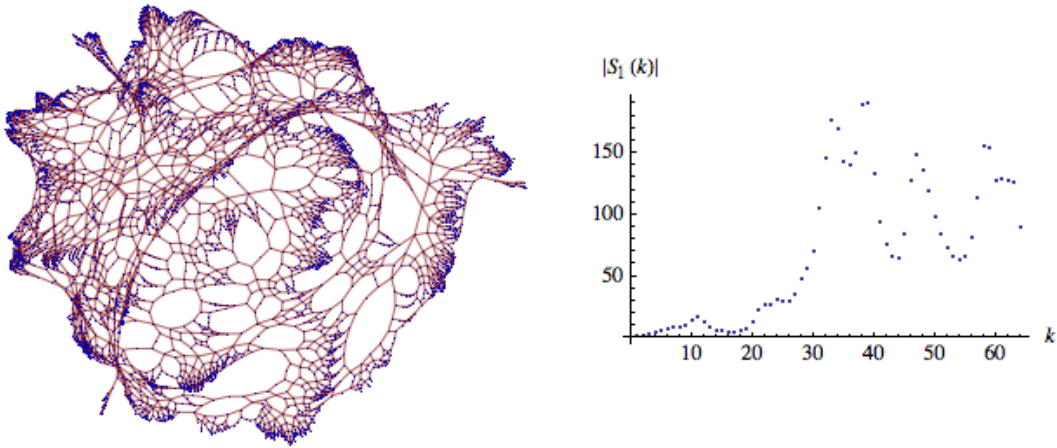

Figure 33 - An example of highly irregular causet from a tag system computation (left), and its node-shell growth (right).

## 3.4 Cyclic tag systems

A *cyclic tag system* is a variant of tag tystems that still manipulates strings on a finite alphabet (e.g. binary). At each step just the first element of the current string is removed, and a block of symbols is appended to its end only if the dropped symbol is a predefined one (e.g. symbol "1"). The attribute 'cyclic' refers to the requirement that blocks to be appended are considered in a cyclic fashion: for example, with a two-step cycle, two blocks $b_0$ and $b_1$ alternate at successive steps (being used only when the dropped symbol is the predefined one). In a system with cycle length $n$, using blocks $b_0$, ..., $b_{n-1}$, if at step $k$ the current string is "1".S, where "." denotes string concatenation and S is a substring, then the rewritten string is S.$b_j$, where $j = k$ mod $n$, otherwise it is simply S.

**Causet construction**

- Every string rewriting becomes an event of the causet.
- A directed edge connects event $e_i$ to event $e_j$ iff $e_j$ drops (reads) a string symbol that was appended (written) by $e_i$.
- A directed edge connects each event $e_i$ to its successor $e_{i+1}$.

The latter link reflects the fact that the procedure keeps memory of the position in the cycle of possible replacements, and every event increments this counter for the next event. This type of link makes the causet totally ordered.

**Proposition 6.** *The causet of a cyclic tag system is planar.*

Proof. Same as for Proposition 5, since that proof is not affected by the cyclic usage of blocks. □

Linear causets, causets with node-shell dimension 2, and causets with negative curvature can be derived from cyclic tag system computations. Two examples are shown in Figure 34, which also includes another example of planar causet with estimated node-shell dimension 3, similar to the causet from the Turing machine of Figure 12, or the one from the string rewrite system of Figure 26. For these three examples, and the next one, computations start with string "1", and cycles have length 3; the replacement blocks are the rows of the 3×3 arrays appearing as labels.

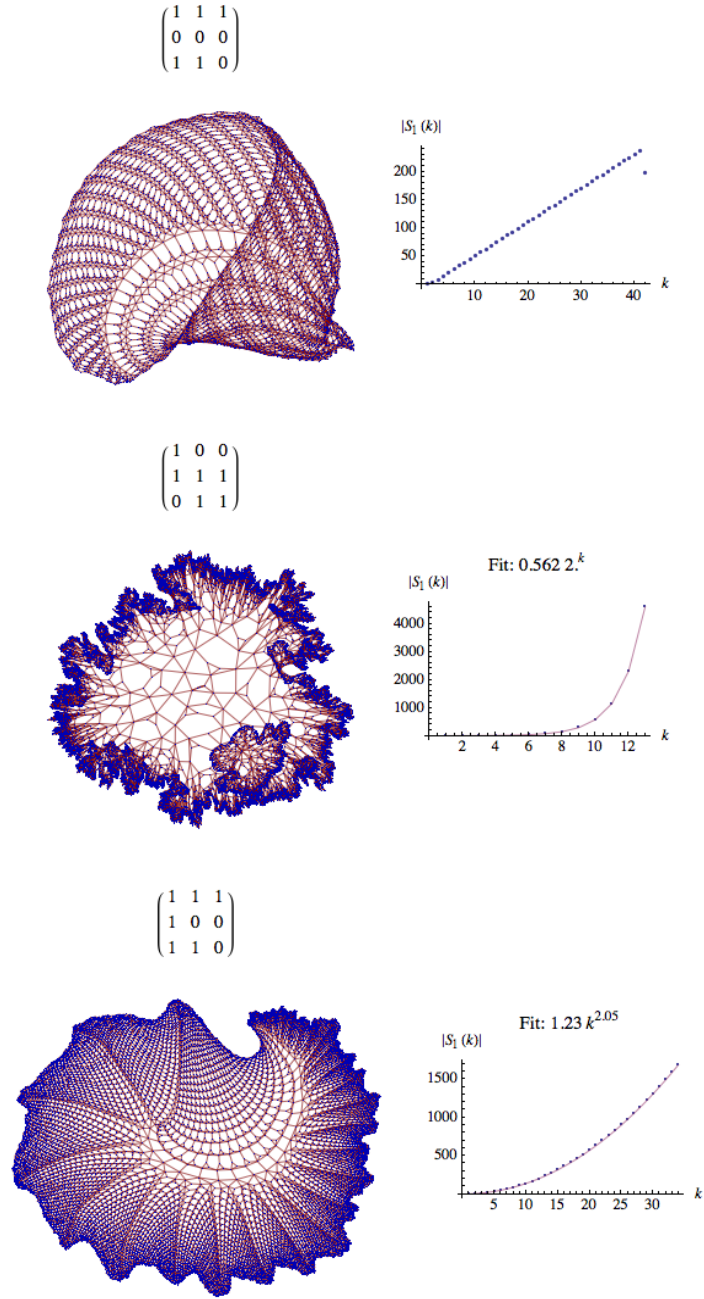

Figure 34 - Causets from cyclic tag systems, with node-shell dimensions 2 (upper), 3 (lower), and with negative curvature (middle).

An interesting causet with negative curvature is also found in this family (see Figure 35), whose node-shell growth reproduces a Fibonacci sequence: more precisely, it satisfies the recurrence $|S_{k+1}(x)| = |S_{k-1}(x)| + |S_k(x)| + 3$, for $k \geq 3$, thus approximating an exponential growth with golden ratio basis.

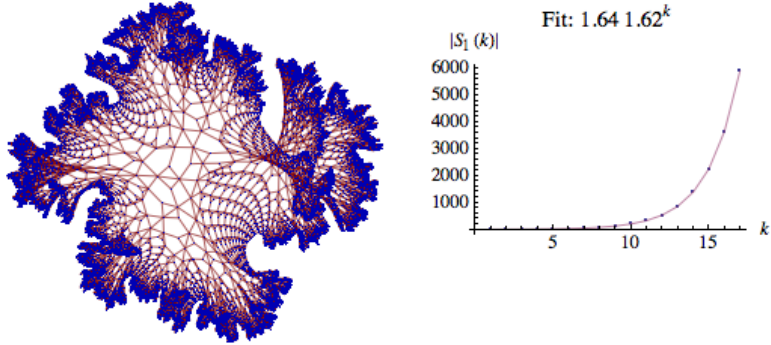

Figure 35 - Causet from Cyclic Tag Systems, with Fibonacci node-shell growth.

# 4. Causets from computations on a higher-dimensional support

With the various models of computation considered in the previous sections we have been able to obtain causets of node-shell dimensions up to 3. In light of the interpretation of a causet as a physical spacetime, we are interested in further exploring the range of dimensions that our technique can attain. Is it possible, for example, to achieve a classical value 4, or values that do *not* closely approximate an integer dimension?

In the sequel we consider two models of computation that operate on two-dimensional supports. One might perhaps conjecture that the dimension of a causet from a given model will be at most one unit higher than the dimension of the underlying support, based on the idea that the latter is a representation of *space* and that building a computation on top of it, yielding *spacetime*-causet, amounts to adding one unit for the time dimension. But in the previous sections we have been able to exhibit computations on 1-dimensional supports -- from Turing machines, string rewrite systems and cyclic tag systems -- that yield 3-dimensional causets, thus falsifying the conjecture. Being aware of this circumstance, let us still consider two models that operate, respectively, on a static, predefined 2-D support, and on a dynamic, flexible one.

## 4.1 Two-dimensional Turing machines

A two-dimensional Turing machine is a Turing machine that operates on a two-dimensional array of cells, rather than on a tape. For each state-symbol pair $(s, a)$, the state transition table will therefore provide a triple $(s', a', d)$, including the new control head state $s'$, the new symbol $a'$ to be written in the current cell $c$, and the move of the control head, indicated as a two-dimensional displacement $d$ from $c$.

**Causet construction**

- Every step of the machine corresponds to an event of the causet.
- A directed edge connects event $e_i$ to event $e_j$ iff $e_j$ reads a cell that was written by $e_i$.
- A directed edge connects each event $e_i$ to its successor $e_{i+1}$.

The last link reflects the causal relation mediated by the udatings of the control head state, as performed by successive events. This type of link makes the causet totally ordered, as already observed for 1-D Turing machines.

The analysis of the causets from 2-D Turing machines with various parameter setting -- alphabet and state-space sizes -- reveals, as usual, an abundance of linear cases, and, to a much lesser extent, of regular 2-D grids and of negatively curved graphs.

In search for more interesting behaviours, we have considered *Turmites* [12], also called *turning machines*. They are a subclass of two-dimensional Turing machines for which the state transition table admits a simplified representation: the step of the control head is described in terms of left- and right-turns relative to the current head orientation. The most famous turmite is Langton's ant [6, 19]. In [13], 44 turmites with particularly interesting behavior have been identified; we have derived causets for all of them, thus discovering a number of quantitative and qualitative features.

As a first qualitative feature, we observe that there is often, but not always, a close similarity between the final TM configuration on the cell array and the shape of the corresponding causet, although the former represents only the final stage of a computation while the latter is a representation of *all* computation steps and of their causal links. Some examples are shown in Figure 36.

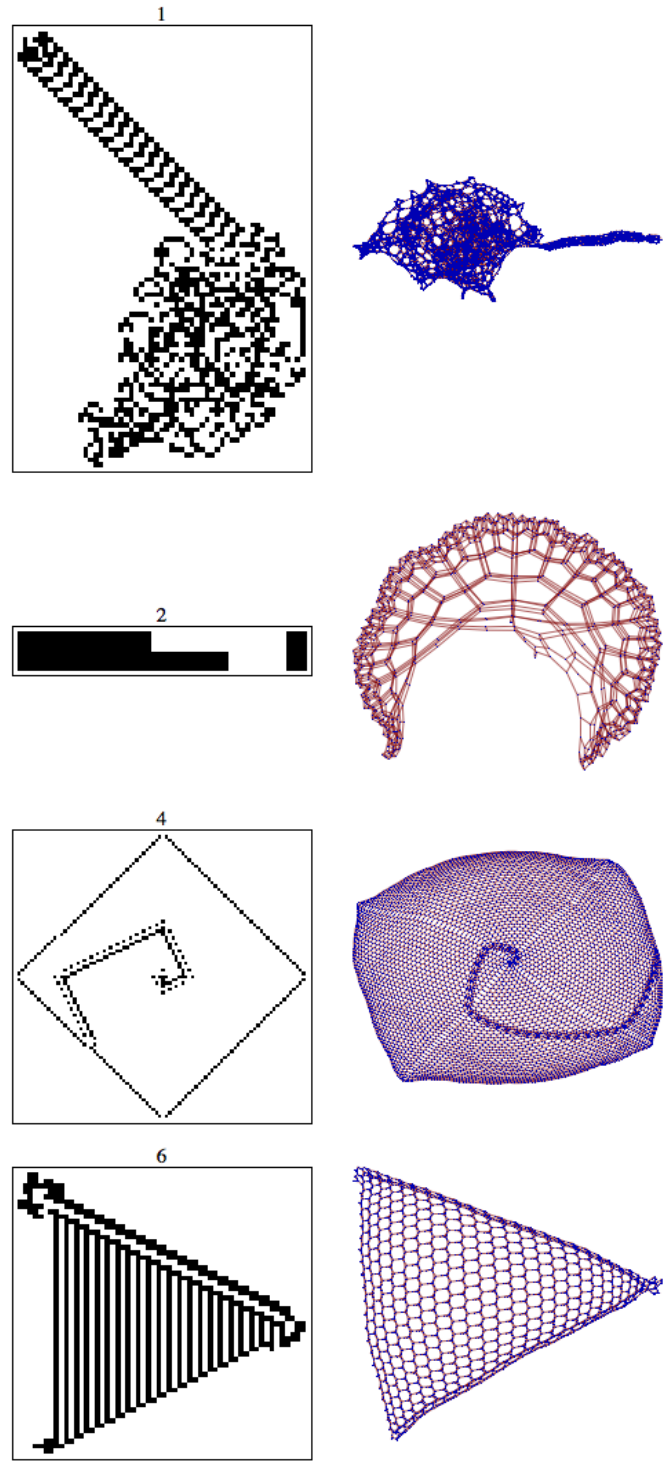

Figure 36 - Final configuration (left) and causet (right) for the computations of some Turmites. Turmites are numbered as in [13].

Similarity between the two structures occurs when the turmite tends not to return to the already visited sites, as it happens, for example, with the 'highway' in Langton's ant (first case in figure). Highways represent themselves an interesting qualitative phenomenon. They are periodic structures that grow forever. In fact, we can extend their definition to encompass generic 'regular' structures, such as the growing triangle of turmite 6, and in this respect they no longer appear as a specific feature of turmites: they have been observed with all previous models, whenever a computation leaves its initial pseudo-random behavior -- which may be very short, or completely missing -- for stabilizing to some regular, visually predictable pattern. One may erroneously conclude that a causet can contain at most one highway. A counterexample is represented by turmite 22 in Figure 37, in which several regular structures spring from a pseudorandom body.

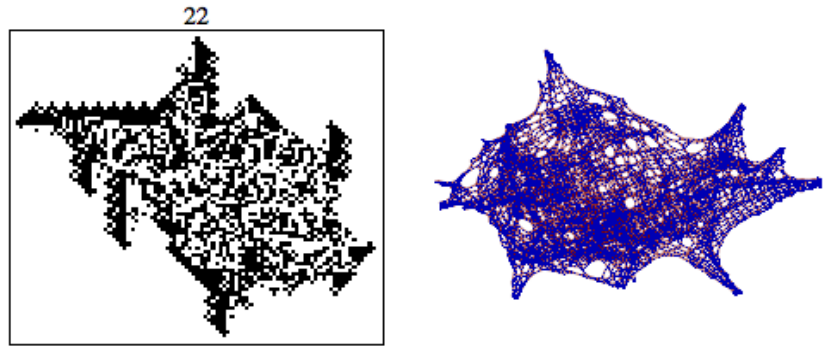

Figure 37 - The turmite numbered 22 in [13] has several highways, which appear both in the final TM configuration and in the causet.

The presence of several highways is explained by the fact that they are built in successive phases. In the simplest case, the control point moves within the highway until it reaches the extreme endpoint of it: but this fact makes it 'bounce back' and reach the pseudorandom body, where it moves for some time until it enters the highway again. (Another multi-highway Turing machine is shown in [21], p. 185, case d.) Thus, the presence of one or more highways does not make the causet unrealistic as a possible physical spacetime, since it does not imply *complete* regularization. In fact, the coexistence of pseudo-randomness and regularity appears as a necessary requirement for realistic models of spacetime.

A final remark on qualitative aspects. One the most attractive features of cellular automata is the emergenge of 'particles', as briefly discussed in subsection 2.2 (Figure 17). In that subsection we only hinted at the possibility to spot particles in causets such as those of Figure 16, but we have now much clearer cases, such as turmite 4 in Figure 36. We did find spiraling localized structures before, e.g. in the top causet of Figure 31, but the case of turmite 4 is somewhat more elaborate. The graph consists of four triangular sectors formed by regular hexagonal grids, separated by four 'radii' formed by sequences of octagons. On top of this structure, a localized perturbation creates a spiral. One could perhaps regard *both* the spiral *and* the 4 radii, formed by octagons, as particles, thus obtaining a new, interesting interpretation of the causet structure as a particle collision diagram.

Note that when these two specific particles collide, they both preserve their structure (in elementary cellular automata, this fact may or may not occur). Note also that the graph is not planar, thus making it impossible to define combinatorial curvature at every node. The non-planarity is introduced by the spiraling particle, which represents, precisely for this reason, a good example of one of the possible particle types discussed in [21].

Incidentally, we find also a turmite whose causet produces a version of the above graph *without* the spiraling particle (turmite 21 in [13]), which turns out to be planar: combinatorial curvature is therefore applicable. In that case, the four grid sectors are flat, since their nodes have combinatorial curvature zero (see Definition 3), but the nodes on the border of the octagons forming the radial particle trajectories have three different non-null curvature values, namely -1/24, -1/12, +1/4. Could this circumstance provide a basis for interpreting the motion of the spiraling particle of turmite 4 in terms of the curvature of its background? This is still an open question.

We believe that the detection of actual particles *in causets from computations* is an original result of this paper. One does detect something similar in turmite diagrams, but there is a clear conceptual difference beween the two cases. The fact that the single, final configuration of a computation retains some trace of a particle's trajectory is accidental (and interesting in itself), but the correct place where to look for particle traces -- analogous to particle worldlines in continuous spacetime -- is the causet.

Let us turn to quantitative features. The maximum node-shell dimension from the root that we find in the considered family of 44 turmites is over 3.5, and is achieved by turmite 11. Figure 38 (right) shows the plot of the fitting non integer monomial $a*x^b$ for the node-shell size data from a 100,000-step computation. The exponent $b = 2.62$ provides a dimension 3.62 estimate. Some indication of the reliability of this estimate is obtained by noting that, when fitting prefixes of growing length of the available node-shell sizes, fluctuations of $a$ and $b$ tend to stabilize for lengths larger than 25.

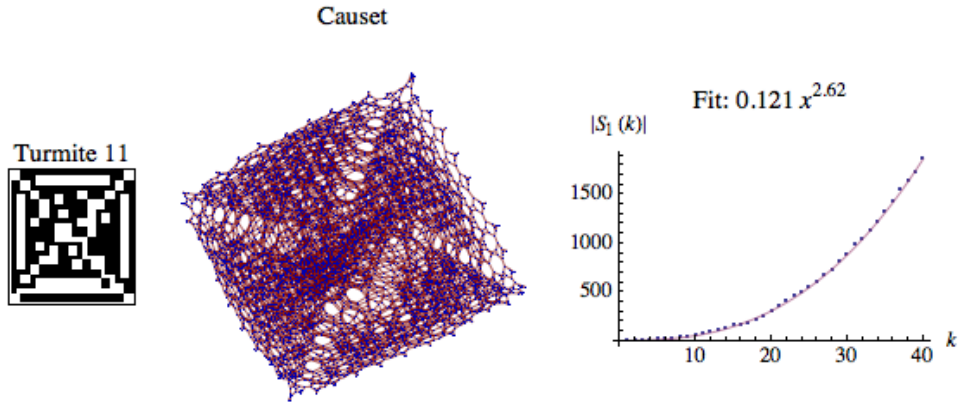

Figure 38 - Final configuration of a 6000-step computation of turmite 11 (left); causet for the same computation (center); node-shell growth and causet dimension estimation, based on a 100,000-step computation (right).

The above example is remarkable in two ways: for the first time, with 2-D Turing machines we obtain a node-shell dimension higher than 3, and one of non-integer value.

## 4.2  Mobile automata on trivalent planar graphs

The last model of computation that we consider is *trinet mobile automata*, one which has been introduced and shortly discussed in [21], and which we have further studied in [1, 2]. A *trinet* is a planar, undirected, *trivalent* graph, that is, one in which all vertices have degree 3. A *trinet mobile automaton* is similar to a mobile automaton on strings (see Subsection 3.1), or a 2-D Turing machine, except that the *stateless* control head moves on a trinet rather than on a tape or a 2-D array of cells. We can imagine the control head positioned on a trinet edge, and oriented toward one of its endpoints. At each computation step the control head does two things: (i) it performs a local graph rewriting, and (ii) it moves to some nearby location. Both the rewrite rule and the next location are chosen by some specified deterministic criterion. In our investigations we have always used the trinet rewrite rules called *Expand* and *Exchange* in Figure 39.

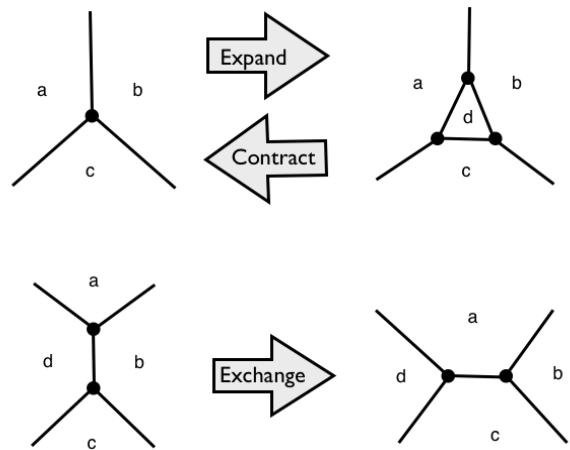

Figure 39 - Rewrite rules for planar trivalent graphs

In some proposals for quantum gravity, the three rules in Figure 39 and their higher dimensional versions (Pachner moves) are used for animating discrete models of spacetime [9, 17]. An important property of these rules is that they form a *complete* set: any trinet can be transformed into any other trinet by their application (the proof is trivial). Note that all three rules affect 3 or 4 faces of the planar graph.

**Causet construction**

- Every step of the machine corresponds to an event of the causet.
- A directed edge connects event $e_i$ to event $e_j$ iff the rewriting performed by $e_j$ affects a planar face of the trinet that has been previously affected by $e_i$ (without further modifications in between).

In [1, 2] we have explored two variants of trinet mobile automata, in both of which rule *Contract* is not used. A first variant, qualified as *three-connectivity preserving*, is defined as follows.

(i) Initial condition: the computation starts with a trinet consisting of two nodes connected by three parallel arcs. (This is the smallest possible three-connected graph. A connected graph is *n-connected* when *n* is the smallest number of edges one has to remove in order to disconnect it.)

(ii) Graph rewrite rule choice criterion: choose rule *Exchange* whenever it does not violate the three-connectivity of the trinet, otherwise choose *Expand*.

(iii) Control head move: move the control head to a new nearby location which depends on the applied rule and on a fixed parameter associated to it (see [2] for a details).

The second variant, qualified as *threshold-based*, is obtained by modifying (ii) as follows: choose rule *Exchange* whenever it does not create trinet faces with less than *k* sides (this being a fixed threshold parameter for each algorithm instance), otherwise choose *Expand*.

The two-parameter policy for control head moves is the same in the two variants, and involves short steps for all parameter settings (following the short-step policy of Turing machines). As a consequence, the faces affected by a step always partially overlap with those affected by the next step, so that events $e_i$ and $e_{i+1}$ are always causally related, and the causet is totally ordered, as in Turing machines. However, due to the above defined construction rule, each node has 3 or 4 incoming, and 3 or 4 outgoing arcs, except for the first node and those at the growth boundary, so that causet structures can still be far from trivial.

As the reader may at this point expect, even with trinet mobile automata computations the large majority of causets are 1-D, linear graphs of no interest. In Figure 40 we illustrate the trinets obtained at the final step of some computations of the three-connectivity preserving variant, and the associated causets: we have a 2-D case, a negatively curved one, and one with a 1-D highway.

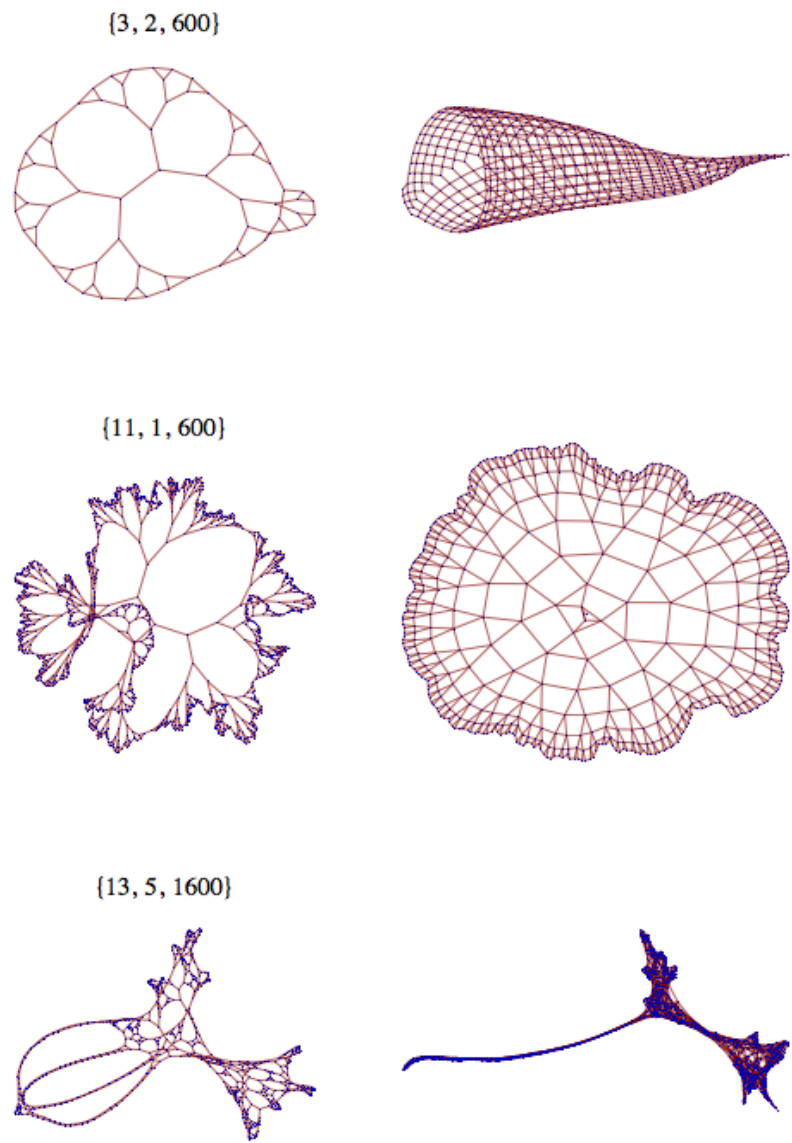

Figure 40 - Final trinet configuration (left) and causet (right) for the computations of some trinet mobile automata. Numeric parameters are as defined in [1, 2].

With trinet mobile automata, causet planarity is no longer guaranteed, and an example of this is provided in Figure 41. The upper-left diagram is a 'revisit-indicator' (see [1]) showing how the control head moves and returns to trinet locations visited earlier. In this case, the diagram yields a version of the numeric sequence known as the *fractal sequence*. (The non-planarity of the causet was tested by the *PlanarQ* function of the *Mathematica Combinatorica* package.) The lower-right diagram shows the node-shell sizes and, in the smaller plot, the ratios $|S_1(k+1)|/|S_1(k)|$ of adjacent values: the flat region reveals the exponential character of the node-shell growth.

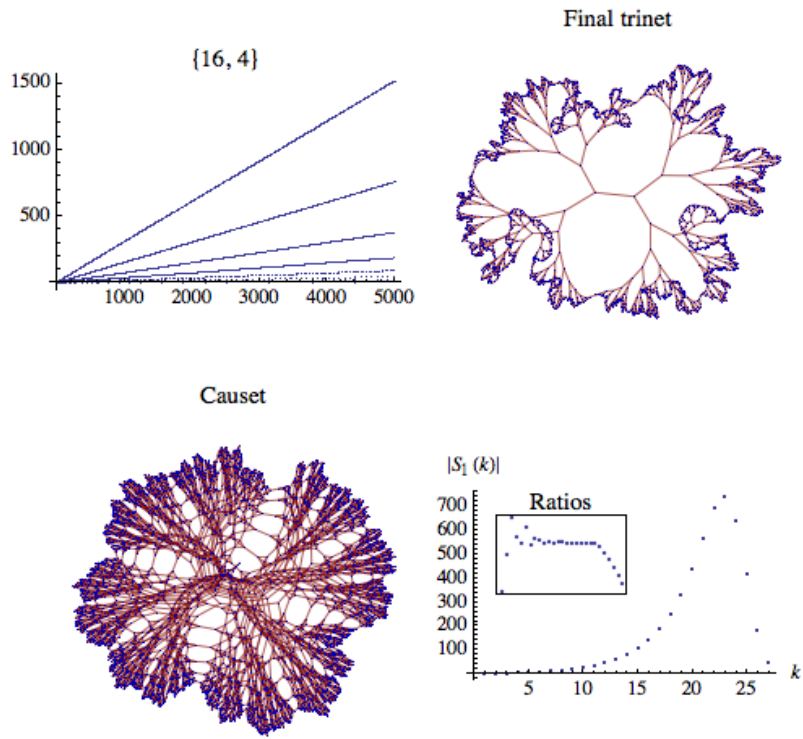

Figure 41 - Final trinet configuration (upper-right), causet (lower-left), and dimensional analysis (lower-right) for the computations of a trinet mobile automaton. The upper-left diagram is a 'revisit-indicator' (see [1]) showing how the control head moves and returns to trinet locations visited earlier in the 5000-step computation.

The threshold-based mobile automaton variant is based on three parameters -- two of them control the head moves and one is the *k* threshold -- thus yielding a larger space of automata. Out of over a thousand parameter settings we have explored (see [1] for details), there are only very few cases that deserve attention with respect to causet dimensionality. We single out two of them.

The first is shown in Figure 42. It is the only causet we have found in this model whose node-shell dimension from the root closely approximates value 3. (Incidentally, node-shell growth in the trinet itself, starting from the oldest node, turns out to be perfectly linear.)

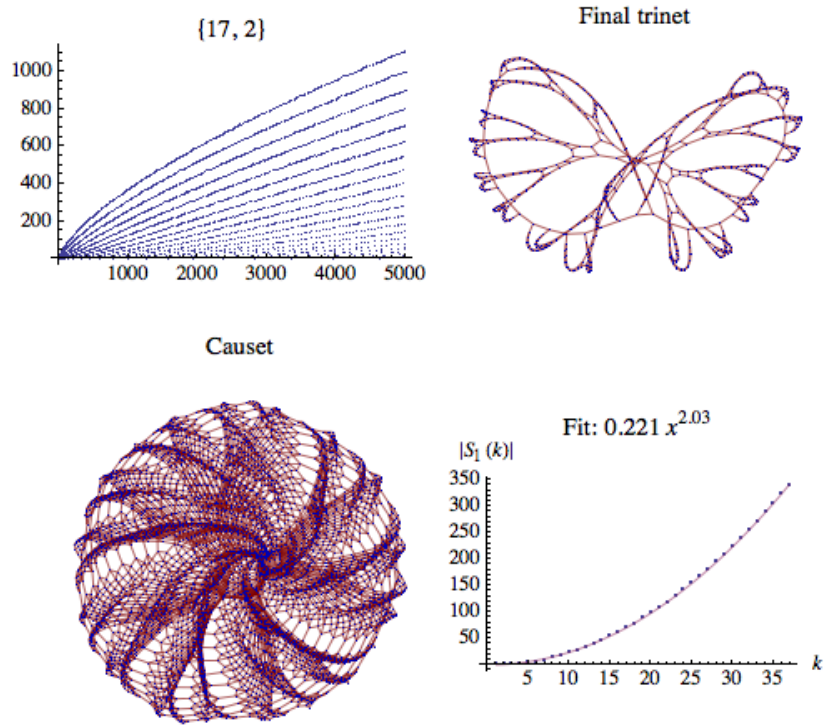

Figure 42 - 'Revisit-indicator' (upper-left - see [1]), final trinet configuration (upper-right), causet (lower-left), and dimensional analysis (lower-right) for a 5000-step computation of a threshold-based trinet mobile automaton (threshold = 6).

The second interesting case is shown in Figure 43.

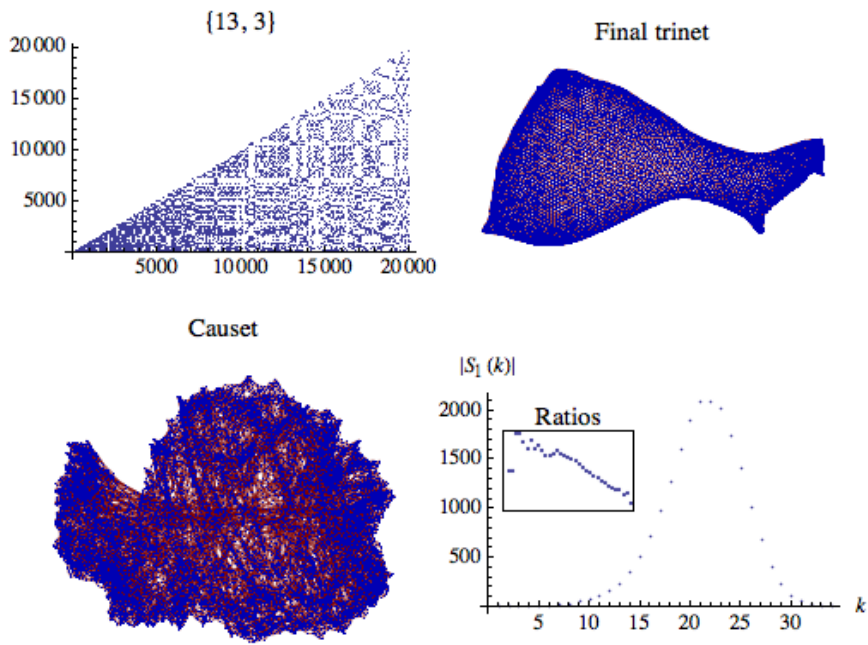

Figure 43 - Final trinet configuration (upper-right), causet (lower-left), and dimensional analysis (lower-right) for the computations of a trinet mobile automaton. The upper-left diagram is the 'revisit-indicator' (see [1]) for this 20,000-step computation.

The trinet itself (upper right corner in Figure 43) had already appeared as unique in its kind. It is a quasi-regular hexagonal grid with a number of non-hexagonal faces, which 'uniformly embeds' on a sphere, in the sense that it does not exhibit the very large 'external' face observed in the planar embeddings of several other trinets produced by this model. However, the pseudo-randomness of the computation is best revealed by the revisit indicator (upper-left plot in figure), showing how the control head keeps covering somewhat uniformly the growing set of nodes and edges of the trinet. The analysis of the causet reveals now a bell-shaped plot for node-shell sizes; this is shown in the lower right plot of Figure 43, with the ratios $|S_1(k+1)|/|S_1(k)|$ of adjacent values shown in the inset plot. Note that, for the bell-shaped normalized gaussian distribution function $f(x) = \frac{1}{\sqrt{2\pi}} e^{\frac{-x^2}{2}}$, the continuous equivalent of the above ratios plot is indeed a linear function with negative slope: $(f(x) + f'(x))/f(x) = 1 - x$.

Finally, three trinet mobile automata computations have been singled out in [1] for their surprising pseudorandom character. Let us examine here their causets.

The first case is provided by the three-connectivity preserving automaton variant, and has control step parameters {17, 8}; see Figure 44. The pseudo-randomness of the computation is revealed both by the revisit indicator (lower-left) and by the actual node-shell size data (lower-right). The revisit indicator is also an immediate indication of the growth rate of the trinet (indifferently in terms of nodes or edges, since these are linearly related): the trinet growth is roughy linear. The node-shell growth appears exponential, with fitting parameters indicated in figure. After about 378,000 steps this automaton stabilizes to regular behavior; its highway is shown in [2].

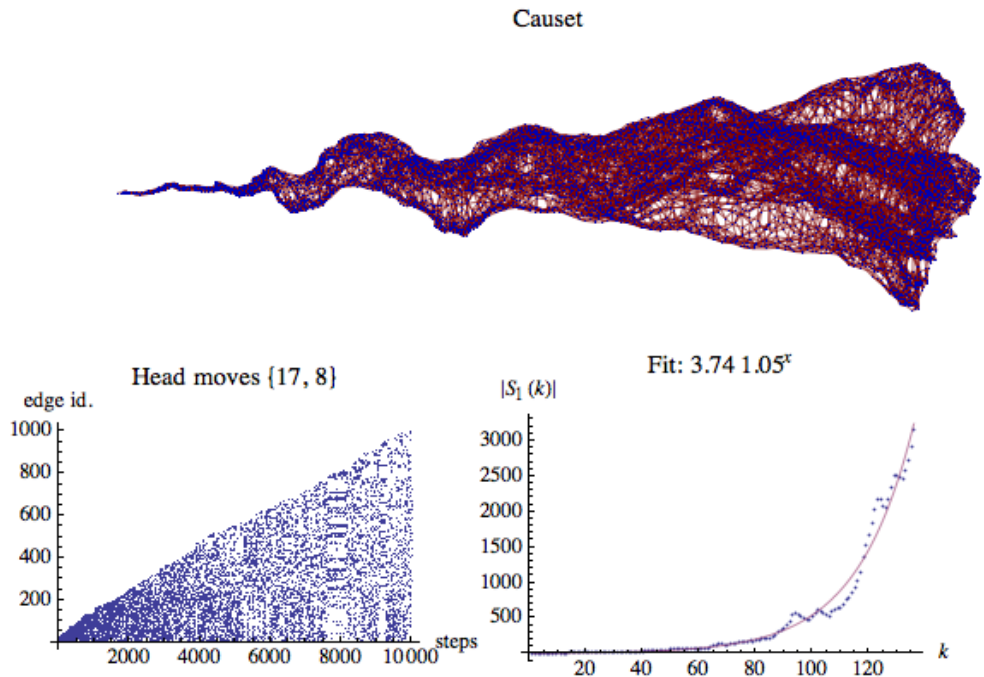

Figure 44 - Causet (upper), and 'revisit-indicator' (lower-left, see [1]) for a 6000-step computation of a trinet mobile automaton (code 17, 8). The dimensional analysis of the causet (lower-right) is carried out for a 100,000-step computation of the same automaton.

The next case is from the threshold-based variant of the algorithm; with threshold 4, and control head moves determined again by parameters {17, 8}, we obtain the computation illustrated in Figure 45. In this case, the revisit indicator still appears pseudo-random, but with an $O(\sqrt{steps})$ growth rate, while the node-shell growth rate of the corresponding causet is close to linear. Unlike the previous case, we have run this computation for a billion steps without finding any sign of regularization.

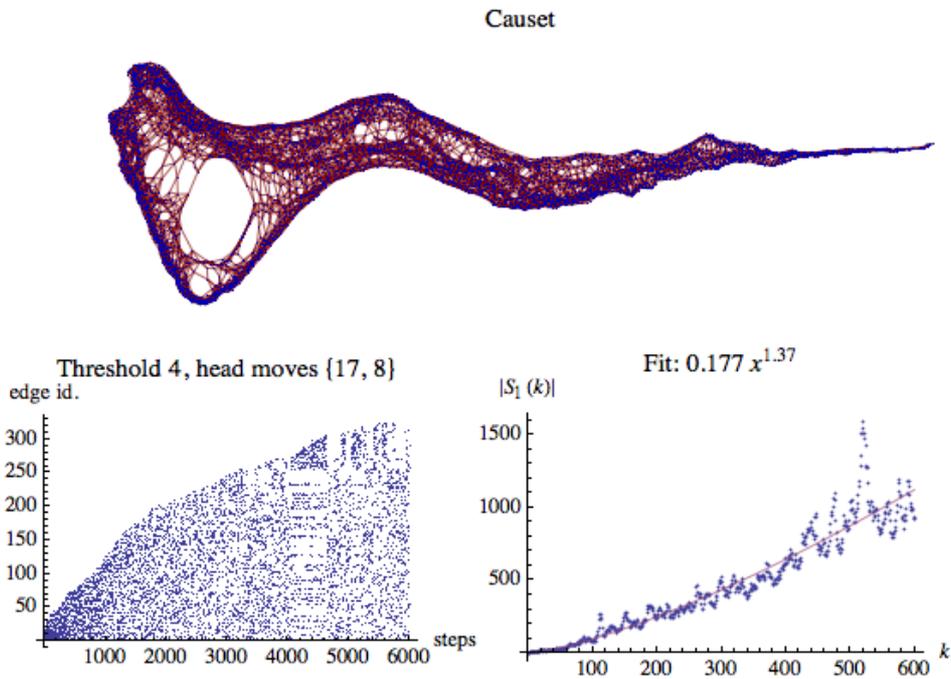

Figure 45 - Causet (upper), and 'revisit-indicator' (lower-left, see [1]) for a 6000-step computation of a trinet mobile automaton (code 4, 17, 8). The dimensional analysis (lower-right) is carried out for a 300,000-step computation of the same automaton.

The third case of pseudorandom computation is similar to the previous one, and is illustrated in Figure 46.

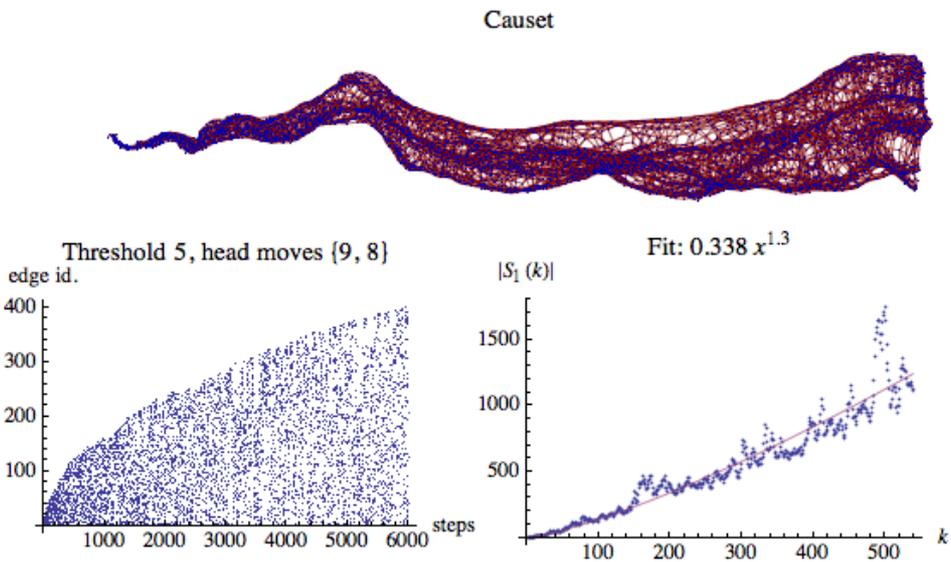

Figure 46 - Causet (upper), and 'revisit-indicator' (lower-left, see [1]) for a 6000-step computation of a trinet mobile automaton (code 5, 9, 8). The dimensional analysis (lower-right) is carried out for a 300,000-step computation of the same automaton.

We conclude this section by mentioning a qualitative property exhibited by one of the above pseudorandom causets, that we consider as very relevant to the interpretation of these structures as a physical spacetime. What type of feature might we expect to see emerge in our computational spacetime models, and in particular in those with pseudorandom character, on larger and larger scales, beside particles and curvature? We suggest that one of the key features would be the *self-organization into components that achieve some form of independence from one another*; this appears as a necessary basis for building further complexity and obtaining a multiplicity of independent phenomena. In Figure 47 we have extracted the final portion of the causet from Figure 45. This segment is formed by 4000 events, that are numbered progressively as the automaton proceeds, and is partitioned into three sub-segments, colored in white, gray, and black. The white and black portions are in direct causal contact, but all gray events have occurred in between. Thus, the 'hole' appearing in the graph is not a pure accident of the specific layout algorithm, but it reflects an actual fork-and-join structure. A trace of the 'hole' is visible also in the revisit indicator at the lower left corner of Figure 45, in the region between 4000 and 5000 steps, which helps understanding this phenomenon as the confinement of the control point into a limited region $R$ of the growing trinet for a relatively large number of steps.

Recall that these causets are totally ordered. Therefore, talking about *causal* independence of the 'detour' would be, strictly speaking, incorrect: we can only say that the gray events have an opportunity to update the elements of $R$ *several times* before letting them act as causality mediators towards the subsequent, black events. Thus, the *internal details* of the white causal substructure are completely unaccessible to the rest of the causet, and this is the precise sense in which we may still talk about an 'independent' causet component.

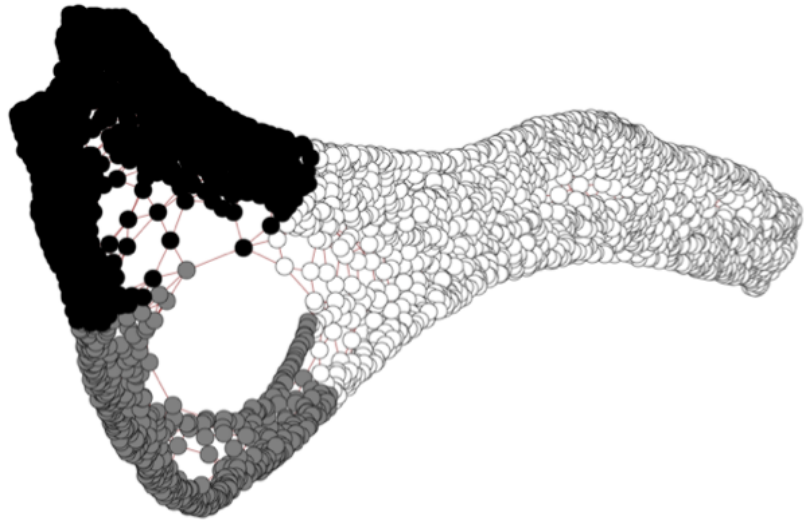

Figure 47 - Final 4000 events of the causet in Figure 45. Events are numbered progressively, as they occur (roughly from right to left), and colored gray up to event 2020, white up to 2649, and then black. Grey and black events are in direct causal contact, while white events create a separate detour.

## 5. Discussion and conclusions

We have investigated a rich family of causal sets -- those that can originate from the 'spontaneous' operation of simple, deterministic formal models of computation -- with the idea to obtain discrete models of physical spacetime, thus offering an alternative to the probabilistic approach [16] adopted in the Causal Set Programme [3], and, at the same time, contributing to the consolidation of a new, causet-oriented viewpoint and experimental effort in the context of the 'Computational Universe Programme'.

Devising a rigorous general criterion for causet construction, and applying it consistently and extensively to a variety of models such as Turing machines, rewrite systems and mobile automata, was one of our goals, and has turned out to be rather straightforward. However, in order to deal with sufficiently complex structures we have decided to keep *all* derived causal arcs, not just the essential 'links' of the transitive reduction; in several cases, the latter operation would completely wash away interesting emergent structures.

The other main goal was two-fold: to observe emergent features of these causets, both quantitative and qualitative, and to identify commonalities among causet classes, as a way to possibly solve the 'tyranny of computational universality' problem posed by Fredkin.

Among *quantitative* emergent features, we have in particular focused on causet *dimension*, using a specific, node-shell dimension estimator, and *curvature*, witnessed by an exponential node-shell growth. In doing so, we have generally restricted to measuring node-shell growth relative to the root. We have proved that all five considered models that operate on a one-dimensional support yield planar causets, a property which is lost when considering higher-dimensional supports. Interestingly, the negatively curved causets that are planar appear in two forms, that correspond to, respectively, the Riemann-Poincaré disk and the Beltrami-Poincaré half-plane models of the hyperbolic plane. We have found examples of causets with node-shell dimension up to 3.5, and cases in which the difference between causet dimension and the dimension of the underlying computation support is larger than 1, thus disproving the conjecture that building a causet out of a computation means increasing at most by one unit -- the time component -- the dimension of the support. An aspect that deserves further investigation is the link between causets and continuous Riemannian manifolds, since concepts such as border, curvature, dimension, space-form, find their natural definition in that setting.

We have also observed the emergence of *qualitative* features such as highways, multiple highways, particles, the mix of regularity and pseudo-randomness, and partially independent macro-regions. Although the idea of obtaining 'particles' out of computations is not new, and is found, for example, in the cellular automata investigated by Zuse [23, p. 64], Conway, and Wolfram, we believe that having observed them precisely in a causal set (the radial and spiraling particles from *turmites*) is a very encouraging result of this paper, since causets represent, in our view, a more genuinely physical representation of spacetime than the rigid grid of cellular automata. The emergence of partially independent macro-regions appears also as a remarkable feature of some *pseudo-random* causets -- ones that are still derived from fully deterministic computations -- and represents a crucial a difference from 'truly random' causets, where this property is definitely not observed.

Several of the computations and corresponding causets shown here are fairly regular. Visually regular graphs are convenient for preliminary explorations of the basic capabilities of the considered model, much in the same way as the 'regular' 1-D cellular automata with random initial conditions -- those in Wolfram class 1 or 2 -- already indicate that localized structures may emerge, without yet showing their interaction. It is clear, however, that the challenge is to spot the emergence of interacting particles, and/or other self-organization phenomena, in less regular cases, such as the pseudo-random causets introduced at the end of the previous section. In the case of *planar* pseudo-random causets, one could proceed by 'coloring' polygonal faces depending on the number of edges, and then searching for possible periodic patterns: some of the causets we have found appear promising under this light. For non-planar graphs, the technique could be extended by generalizing faces to higher-dimensional simplices. This needs further work.

Did we find commonalities among causet classes? The two most obvious common aspects are (i) the appearance of periodic, one-dimensional causets, that represent the vast majority in all models, and (ii) the less frequent appearance of regular (predictable) negatively curved causets, which correspond to *nested* computations, as they are called in [21]. If our physical spacetime matched one of these two forms, the problem of choosing a specific model of computation would perhaps disappear. But the complexity of physical spacetime is obviously much higher, thus we need to focus on the most complex elements in the different causet classes. Elements at that level collectively exhibit a large variety of forms, but, unfortunately, each class seems to possess its own distinctive features, so that the possibility to identify some shared, universal forms appears remote. Typically, causet node degrees do reflect the details of the corresponding computation model: we have seen, for example, that Turing machines yield nodes of degree 4, mobile automata on strings may originate nodes with unbounded degree, and trinet mobile automata yield degrees 7 or 8. These differences have an impact on the type of textures that causets can implement. Notions of causet equivalence abstracting from node-degree details would be useful here.

A more dramatic discriminating factor is that of planarity. We have shown that five of the considered models are unable to produce non-planar causets, while the others generally can; however, *all* of these models are capable Turing-universal computation. Thus, universality does not need to violate causet planarity, but, on the other hand, non-planar causets may offer a richer variety of forms and patterns, and achieve higher dimensions, as our experiments have confirmed.

In conclusion, our analysis has been more successful in discovering variety than unity, and is far from providing evidence that, when looking for a physically meaningful computational spacetime, all Turing-universal models would be equally good: the 'bold' conjecture mentioned in the introduction is far from being confirmed.

Thus, with respect to the 'Computational Universe Programme', we are still left with the problem of choosing among universal models of computation, but the contribution of this paper has been to take an alternative approach to Fredkin's, and to shift the focus from the specific features of those models to those of their associated causal sets. And with respect to the Causal Set Programme, we have proposed a novel, fully deterministic, still classical (as opposed to quantum-mechanical) technique for growing causality structures, which represents an alternative to the probabilistic technique described in [16].

**Acknowledgements**. This work is partly supported by the CNR Project RSTL-XXL. We express our gratitude to Stephen Wolfram, Alex Lamb and Hans-Thomas Elze for various lively discussions on the topics covered in the paper, and to Wolfram Research (www.wolfram.com) for kindly granting permission to use two figures from [21].